\PassOptionsToPackage{pdfpagelabels=false}{hyperref}
\documentclass[useAMS,fleqn,usenatbib]{mnras}
\pdfoutput=1
\usepackage{graphicx}
\usepackage{amsmath,amssymb,amstext,color}
\usepackage[T1]{fontenc}
\usepackage{ae,aecompl}
\usepackage[utf8]{inputenc}
\usepackage{newtxtext,newtxmath}
\usepackage[figure,figure*,table,table*]{hypcap}
\usepackage[dvipsnames]{xcolor}

\input{macros.sty}
\input{hyperlink-year-only-natbib-patch}

\title[Concentration and Halo Merger History]{Concentrations of Dark Haloes Emerge from Their Merger Histories}

\author[K.~Wang et al.]
{Kuan~Wang,$^{1,2}$\thanks{E-mail: \email{kuw8@pitt.edu}}
Yao-Yuan~Mao,$^{3}$\thanks{E-mail: \email{yymao.astro@gmail.com}; NHFP Einstein Fellow}
Andrew~R.~Zentner,$^{1,2}$
Johannes~U.~Lange,$^{4, 5}$
\newauthor
Frank~C.~van den Bosch,$^{6}$
Risa~H.~Wechsler $^{5,7}$
\vspace*{6pt} \\ 
$^{1}$Department of Physics and Astronomy, University of Pittsburgh, Pittsburgh, PA 15260, USA\\
$^{2}$Pittsburgh Particle Physics, Astrophysics, and Cosmology Center (PITT PACC), University of Pittsburgh, Pittsburgh, PA 15260, USA\\
$^{3}$Department of Physics and Astronomy, Rutgers, The State University of New Jersey, Piscataway, NJ 08854, USA\\
$^{4}$Department of Astronomy and Astrophysics, University of California, Santa Cruz, CA 95064, USA\\
$^{5}$Kavli Institute for Particle Astrophysics and Cosmology and Department of Physics, Stanford University, Stanford, CA 94305, USA\\
$^{6}$Department of Astronomy, Yale University, PO. Box 208101, New Haven, CT 06520, USA\\
$^{7}$Kavli Institute for Particle Astrophysics and Cosmology and Department of Particle Physics and Astrophysics, \\
\phantom{$^{7}$}SLAC National Accelerator Laboratory, Menlo Park, CA 94025, USA
\vspace*{-15pt}
}

\pubyear{2020}

\begin{document}
\label{firstpage}
\pagerange{\pageref{firstpage}--\pageref{lastpage}}
\maketitle

\begin{abstract}
The concentration parameter is a key characteristic of a dark matter halo that conveniently connects the halo's present-day structure with its assembly history. 
Using `Dark Sky', a suite of cosmological $N$-body simulations, we investigate how halo concentration evolves with time and emerges from the mass assembly history. 
We also explore the origin of the scatter in the relation between concentration and assembly history.
We show that the evolution of halo concentration has two primary modes: (1) smooth increase due to pseudo-evolution; and (2) intense
responses to physical merger events. 
Merger events induce lasting and substantial changes in halo structures, and we observe a universal response in the concentration parameter.
We argue that merger events are a major contributor to the uncertainty in halo concentration at fixed halo mass and formation time. 
In fact, even haloes that are typically classified as having quiescent formation histories experience multiple minor mergers.
These minor mergers drive small deviations from pseudo-evolution, which cause fluctuations in the concentration parameters and result 
in effectively irreducible scatter in the relation between concentration and assembly history.
Hence, caution should be taken when using present-day halo concentration parameter as a proxy for the halo assembly history, especially if the recent merger history is unknown. 
\end{abstract}

\begin{keywords}
dark matter -- galaxies: haloes -- galaxies: evolution -- methods: numerical
\end{keywords}



\section{Introduction}
In the concordance, $\Lambda$CDM cosmological model 
\citep[e.g.,][]{komatsu_WMAP711,planck13,Planck2018,DES2018}, 
the formation of galaxies and clusters proceeds hierarchically: 
smaller dark matter haloes are the first to collapse and these 
haloes grow larger through mergers. 
Dark matter haloes form around peaks in the initial density field, 
and gas cools and condenses to form galaxies 
within the potential wells provided by these haloes
\citep{whiterees78,blumenthal_etal84}.
The formation and evolution of haloes and 
galaxies are thus inextricably linked.
A key goal of developing a modern theory of 
structure formation has thus been 
to understand the detailed connection between 
galaxy properties and 
the structure and assembly histories of the 
dark matter haloes in which they form.

Contemporary computational hardware and algorithms enable large-volume, 
high-resolution, 
gravity-only, N-body simulations of structure formation, 
as well as the rapid analysis of these 
simulations \citep[e.g.,][]{springel_etal05,bolshoi_11,bolplanck2016,klypin_etal16,potter_etal2017,heitmann_2019_outerrim,deRose2019_aemulus}. 
Consequently, simulations have largely replaced analytic models 
\citep[e.g.,][]{pressschechter74,Bond91,seljak00,berlind02,cooray02}
as the primary framework for the interpretation of large-scale structure measurements. 
In these analyses, dark matter haloes are considered the basic units of 
nonlinear structure and observable statistics are computed by 
associating galaxies with haloes using some physically-motivated, empirical model \citep[see][for a recent review]{wechsler_tinker18}.
Therefore, an understanding of halo structure is necessary in order to 
interpret observations and to test models of galaxy formation, cosmology, 
and/or the nature of the dark matter.

The most commonly accepted model for the density profiles of haloes 
is the two-parameter profile defined by Navarro, Frenk, and White 
\citep[][NFW hereafter]{nfw95,nfw96,nfw97},
\begin{equation}
  \rho_{\rm NFW}(r)=\frac{\rhos}{(r/\rs)(1+r/\rs)^2},  
\end{equation}
where $\rhos$ is the inner scale density, 
and $\rs$ is the scale radius, 
which characterises the transition from 
$\rho(r) \propto r^{-1}$ in the inner halo to 
$\rho(r) \propto r^{-3}$ in the outer halo.
Though refinements to the NFW profile have been 
suggested \citep[e.g.,][]{moore98_profile,einasto65,gao08,navarro10}, 
the NFW profile successfully describes the 
general structure of haloes found in simulations 
and has become the \textit{de facto} standard 
halo profile.

It is now customary to quantify the relative concentration of a 
halo's mass toward its centre using the \textit{concentration parameter}: 
\begin{equation}
    \cvir=\Rvir/\rs,
\end{equation}
where $\Rvir$ is the halo's virial radius.
NFW discovered that the concentration 
parameter is a decreasing function of halo mass. 
This is known as the concentration--mass relation, 
which has since been extensively studied \citep[e.g.,][]{bullock01,wechsler02, maccio08,prada2012,ludlow2014,Correa2015III,diemer2015,klypin_etal16,child2018}.

In addition to establishing the {\em de facto} standard 
density profile, NFW suggested a relationship 
between halo concentrations and halo mass assembly 
histories, and this was quickly seized upon in 
subsequent studies.
For example, \citet{salvador-sole1998} 
argued that violent relaxation, induced by the rapidly-fluctuating 
gravitational potentials present during halo mergers, 
rearranges halo structure leading to a nearly universal 
mass profile. 
Based on the framework first proposed by \citet{nfw97}, \citet{bullock01} quantitatively modelled halo concentration by relating it with an epoch of initial halo collapse that sets the initial inner halo density. 
\citet[][W02]{wechsler02} found a general functional form of the mass 
assembly history \citep[see also][]{vdb02,tasitsiomi04,McBride2009,Wu2013,vdBosch14, Correa2015I}, and established a tight correlation between halo 
concentration and the characteristic formation epoch, 
$a_c$, at which ${\rm d}\log M/{\rm d}\log a$ 
falls below a specified value of $S$ 
(W02 took $S=4.1$ for their primary results). 
Later works found that, on average, 
the halo mass assembly history can be
roughly divided into an early phase of fast accretion 
that builds up the potential well, and a late phase of slow accretion 
that adds mass without significantly changing the potential well 
\citep[e.g.,][]{zhao03,li_mo_vdb_07,zhao_etal09}.
In this scenario, the fast accretion phase sets an approximately 
universal initial concentration, 
while the concentration only grows slowly 
during the slow accretion phase. 
Moving beyond the one-parameter description 
characterised by the concentration parameter, 
\citet{ludlow_etal13} studied the entire halo mass profile, 
and interpreted it in terms of the entire halo assembly history, demonstrating a link between the two.

The physical nature of halo mass growth 
was further studied by
\citet{diemer_etal13}, who distinguished ``physical evolution'' from ``pseudo-evolution," which refers to the increase in halo mass 
resulting from the dilution of the background density rather than the 
coherent infall of matter associated with mergers.
The virial radius of a halo, $\Rvir$, 
is typically defined as the radius of the spherical 
region within which the average density is some multiple 
(the exact value depends upon the specific analysis) 
of the mean density or critical density of the universe. 
As the universe 
expands and the reference density dilutes, 
halo radii and halo masses grow even in 
the absence of any physical mass accretion onto the halo. Pseudo-evolution 
increases halo radii, so it also proportionally increases halo concentrations. 
In the majority of models proposed to explain the relation between concentration and mass, 
and/or the relation between concentration and formation time, the scale radii of haloes 
were assumed to be set during an initial stage of rapid mass acquisition.
After this initial phase, scale radii were typically assumed to be fixed 
or to evolve only slowly. In these models, concentrations subsequently increase 
as haloes slowly acquire mass via mergers, smooth accretion\footnote{In the present work we consider smooth physical accretion as the limit of minor mergers and do not treat it separately.}, or pseudo-evolution, 
all of which increase $\Rvir$ while $\rs$ is assumed to remain approximately 
fixed. 
Differentiating between mass growth modes has had an important role in interpreting the evolution of the concentration.
\citet{wang2011}, for instance, separated mergers that affect inner regions of haloes from ``diffuse'' accretion during which the inner regions remain stable; this later effect in fact includes pseudo-evolution.
However, the assumption of a stable inner region and constant scale radius would only hold if the halo has a perfectly quiescent assembly history.
\citet{li_mo_vdb_07}, for example, found that the slow accretion phase is still dominated by minor mergers, which, as we will show, can impact the scale radius.

The scatter around the mean relation between concentrations and mass assembly histories, and the origin of such scatter, have also been of considerable interest.
W02 demonstrated that a large part of the scatter in the concentration--mass relation can be attributed to different 
formation times at a fixed mass, but the remaining scatter in the relation of concentration and formation time prompts further investigation.
W02 and \citet{maccio08} found that the scatter in the concentration is reduced
when the haloes with recent mergers are excluded from the sample, 
which suggests that mergers contribute to this scatter.
\citet{ludlow2012} found that haloes identified when they are substantially out of equilibrium, primarily due to mergers, experience oscillations in their concentrations. \citet{lee2018} observed similar behaviour in their phase-space analyses of haloes during post-merger relaxation.
This could result in a scatter in the concentration, depending on the time of measurement.
It is also natural to expect that, beyond the identification 
of a single proxy for the formation time of a halo, 
the various details of mass assembly histories 
play a part in shaping halo structure.
\citet{neto07}, for instance, found evidence suggesting 
that halo concentration depends not only on the mass assembly 
history of the halo, but also on the mass assembly histories of 
the haloes that merged to form the final halo.
A more recent study by \citet{Rey_etal2019} demonstrated that halo concentrations are sensitive to both the smoothness of the merger history and the order in which mergers happen, by generating versions of the same halo with different assembly histories \citep[see also][]{Roth2016_GMhalo}.
\citet{Johnson2020} developed a model for predicting scale radii and hence concentrations, which takes into account the entire structure of the merger tree, and were able to better capture the scatter than previous models \citep[e.g.,][]{ludlow2016,benson2019}.

In this study, we seek a detailed understanding of the relationship 
between the mass assembly histories of haloes and their 
concentrations.
We perform a systematic search to identify characteristics 
of the mass assembly history that can effectively predict 
present-day concentration. 
Various summary statistics of the mass assembly history are 
highly correlated with the present-day concentration. In this work, we explore different ways to represent the mass assembly history to further optimise such correlations.
We then study the evolution of the concentration parameter, and investigate how pseudo-evolution and merger events impact the evolution of concentration, 
both for individual haloes and statistical samples. 
We study how mergers contribute to the scatter in the relation between concentration and mass assembly history.

This paper is organised as follows.
In \autoref{sec:sims}, we describe the simulations we use and specify the selection criteria for our samples.
In \autoref{sec:c-MAH}, we report the correlation between halo concentration and mass assembly history that we find in our samples.
The separate roles of pseudo-evolution and physical growth in the evolution of halo concentration and halo scale radius are examined in \autoref{sec:pseudo-phys}.
We discuss our findings and draw conclusions in \autoref{sec:summary}.

\section{Simulation and Sample Selection}
\label{sec:sims}

\subsection{Simulation}
\label{sec:darksky}

In this work, we use the Dark Sky Simulations, a suite of cosmological, gravity-only simulations \citep{skillman2014}.
The Dark Sky Simulations are run with the \textsc{2HOT} code \citep{warren2013_2HOT}, adopting a flat cosmology with $h=0.688$, $\Omega_{\rm m}=0.295$, $n_s=0.968$, and $\sigma_8=0.834$.
We use two of the Dark Sky Simulations: \texttt{ds14\_b} and \texttt{ds14\_i}.
The \texttt{ds14\_b} box has a volume of $(1\,\Gpch)^3$, with $10240^3$ particles; however, the halo catalogues and merger trees that we use are generated with a downsampled version\footnote{Unfortunately, halo catalogues and merger trees of the full \texttt{ds14\_b} simulation are not available due to computational infeasibility.} of \texttt{ds14\_b} that has only $10240^3/32 \simeq 3225^3$ particles, with an effective mass resolution of $2.44 \times 10^9$\,\Msunh.
The \texttt{ds14\_i} box has a volume of $(400\,\Mpch)^3$, 
with $4096^3$ particles, and hence a mass resolution of 
$7.63 \times 10^7\,\Msunh$.

Both simulations have outputs at 99 epochs,
$a=\left \{0.06,0.065,...,0.09,0.095,0.1,0.11,0.12,...,0.99,1\right \}$. 
The halo catalogues are generated by the \textsc{Rockstar} halo finder 
\citep{rockstar}, using the virial definition as the halo boundary, 
corresponding to a spherical overdensity of $\subt{\Delta}{crit}$, which 
takes the value of 100.46 at $a=1$ in this cosmology, with respect to the 
critical density \citep{bryan_norman98}.
Throughout the paper, we use $\Mvir$ as the halo mass and $\cvir$ as the 
halo concentration, and we will omit the subscript ``vir'' in places for 
brevity.
 
\textsc{Rockstar} identifies haloes in the six-dimensional 
phase-space, utilising both position and velocity information.
This algorithm greatly improves performance in distinguishing subhaloes 
and tracking merger events, compared with friends-of-friends algorithms that are 
based solely on dark matter particle positions.
Subhaloes are haloes with centres that fall within the the virial radius of a larger halo, while haloes with centres that do not lie within the virial radius of any larger halo are referred to as host haloes.
In \textsc{Rockstar}, the scale radius, $\rs$, is directly fitted using a $\chi^2$ fit of the NFW profile.
The particles associated with a halo are divided into up to 50 radial
equal-mass bins, with a minimum of 15 particles per bin, and radial bins that are smaller than 3 times the force resolution scale are assigned a low weight in the estimation of $\chi^2$, to suppress resolution effects at small scales.
The concentration is calculated from $c = R/\rs$, where $R$ is the halo radius. 
As most halo finders do, \textsc{Rockstar} fits the radially averaged profile, and includes substructures in the fit. It is reasonable to expect that the results of our analyses would be different if substructures were removed from the profile. A lower bound is enforced on the fitted concentration, $c\geq 1$.
We have tested that our conclusions do not rely on the fitting scheme, and are not affected qualitatively when $V_{\rm max}/V_{\rm vir}$ is used as a proxy for concentration \citep{klypin11_rs}.

The merger history is analysed using the 
\textsc{Consistent Trees} merger tree builder \citep{behroozi_trees13}. 
At each merger event, we refer to the merging halo 
that shares the most particles with the resulting halo,
as the {\it main progenitor halo.}
Merger trees are constructed by tracing the evolution of a 
halo from today backward in time. 
The {\it main branch} 
of the halo merger tree follows 
the main progenitor halo at each merger event.
We refer the interested reader to \citet{rockstar} and \citet{behroozi_trees13} for details.

\subsection{Sample Selection}
\label{sec:sample}

\subsubsection{Present-day Mass Samples}
\label{sec:mass_sample}

We first study three host halo samples defined by present-day virial mass, 
around $10^{12}\Msunh$, $10^{13}\Msunh$, and $10^{14}\Msunh$ respectively.
The details of the selection are listed in \autoref{tab:mass_sample}.
We choose to select the halo samples from different simulation boxes because 
the mass resolution of the 1$\Gpch$ simulation does not suffice to resolve 
the internal structures of lower-mass haloes at early times, while the number 
of cluster-size haloes in the 400$\Mpch$ simulation is relatively limited.
Hereafter we will refer to these three samples as the 
$10^{12}\Msunh$, $10^{13}\Msunh$, and $10^{14}\Msunh$ samples.

\begin{table}
    \centering
    \begin{tabular}{c|c|c|c}
    \multicolumn{4}{|c|}{\textbf{Present-day Mass Samples}} \\
    \hline\hline
        Box & $M_{\rm{min}}$ & $M_{\rm{max}}$ & Sample size \\
    \hline
        400\,$\Mpch$ & $10^{12}\Msunh$ & $1.1\times10^{12}\Msunh$ & 21099 \\
        400\,$\Mpch$ & $10^{13}\Msunh$ & $2.0\times10^{13}\Msunh$ & 14543 \\
        1\,$\Gpch$   & $10^{14}\Msunh$ & $3.8\times10^{15}\Msunh$ & 25438 \\
    \hline\hline
    \end{tabular}
    \caption{Three present-day mass halo samples.
    We list the simulation box from which each sample is selected, the lower and upper bounds of virial mass, and the resulting sample sizes.}
    \label{tab:mass_sample}
\end{table}

\subsubsection{Major Merger Samples}
\label{sec:amm_sample}

To examine major merger events and the 
impacts of these mergers on halo structure, 
we identify the haloes that undergo major mergers in their main branches
at the time step preceding $a=0.33,0.50,0.67$ and $0.80$, corresponding to 
redshifts of $z=2,1,0.5$ and $0.25$ respectively\footnote{The time of merger is defined as the first snapshot in which the centre of the smaller progenitor has entered the virial boundary of the main progenitor and the smaller progenitor has become a subhalo.}. Our sample selection is based on the major mergers identified by \textsc{Consistent Trees}, which defines major mergers as mergers in which the ratio of the masses of 
the progenitors exceeds $1/3$.
We compare our major merger samples with a 
control group of haloes selected randomly 
from the simulation. 
For all these samples, we require the haloes to be host haloes today and at the time of the 
major merger. We further require each halo in our samples to have mass above $4\times10^{10} h^{-1}\text{M}_\odot$ since $z=2$ to circumvent the effect of mass resolution\footnote{We have tested that this resolution requirement does not affect our results.}.
All the samples are selected from the 400\,$\Mpch$ box.  
Each halo can belong to multiple samples; a halo that undergoes major mergers at more than 
one snapshot of interest will be included in multiple major merger samples, 
and the random sample can include haloes 
that are in the major merger samples.
The size of each sample is listed in \autoref{tab:amm_sample}.

\begin{table}
    \centering
    \begin{tabular}{c|c}
    \multicolumn{2}{|c|}{\textbf{Major Merger Samples}} \\
    \hline\hline
        Sample & Sample size \\
        \hline
        $a_{\rm MM}=0.33$ & 58241 \\
        $a_{\rm MM}=0.50$ & 17784 \\
        $a_{\rm MM}=0.67$ & 10426 \\
        $a_{\rm MM}=0.80$ & 7091  \\
        Random             & 95087 \\
        \hline\hline
    \end{tabular}
    \caption{Sample size of each major merger sample and the random sample.
    The parameter $a_{\rm MM}$ denotes the scale factor of the universe when the major merger occurred.
    The samples are not mutually exclusive.}
    \label{tab:amm_sample}
\end{table}

\section{Relation between concentration and mass assembly history}
\label{sec:c-MAH}

In this section, we revisit the connection between halo concentration 
and halo mass assembly history using the Dark Sky Simulations, 
exploring several aspects of halo mass assembly histories. 
In all cases, 
we study samples of haloes within a narrow range of contemporary 
mass and further control for any mass-dependent effects within 
each sample. This implies that these results also characterise the 
correlations between present-day scale radii and halo mass assembly 
histories because haloes of fixed mass have identical virial radii.

\begin{figure}
    \centering
    \includegraphics[trim=0 20 20 20, clip, scale=0.43]{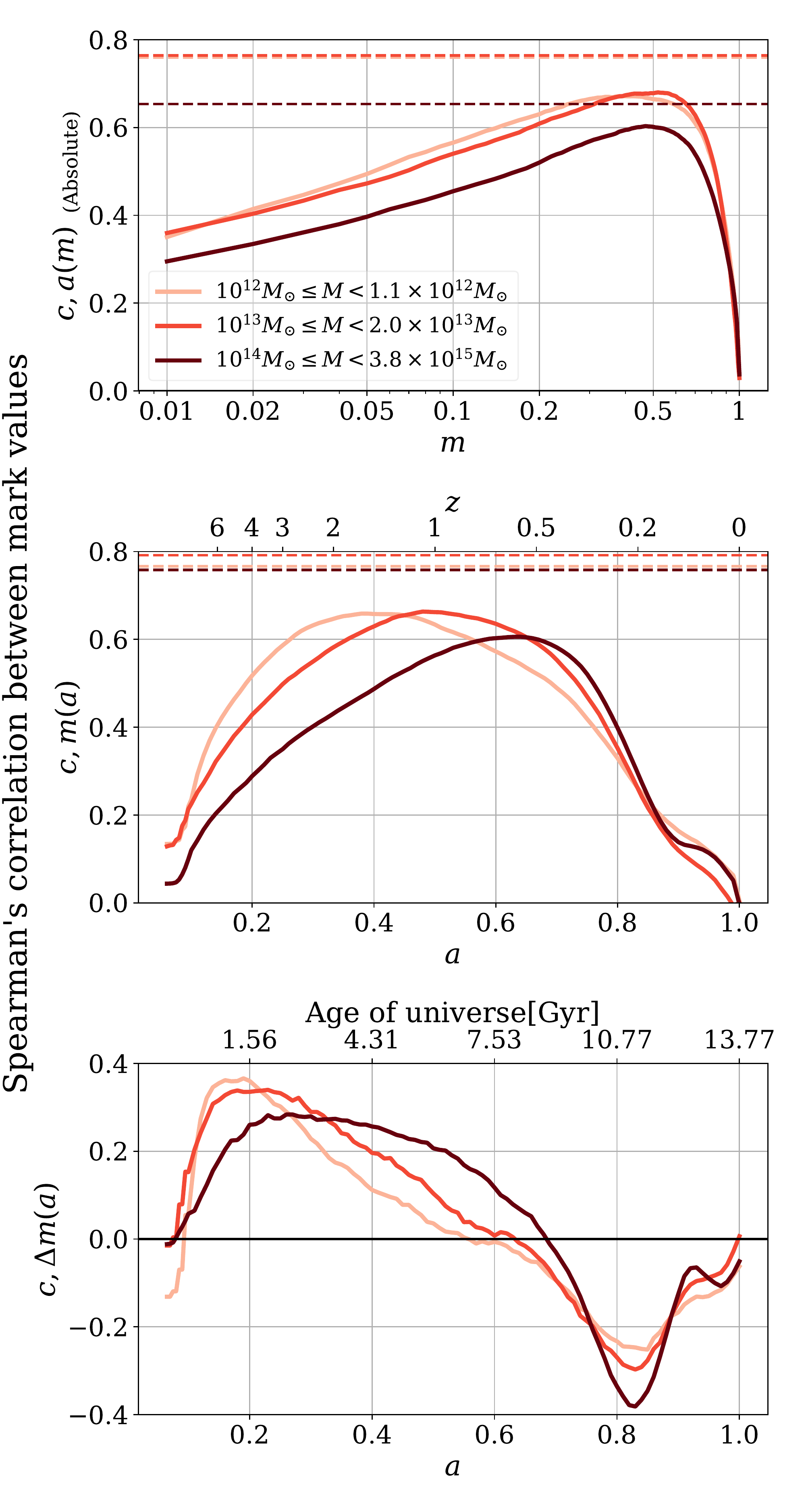}
    \caption{Spearman's correlation between the mark values of haloes' present-day concentrations and mass assembly histories.
    In the top panel, mass assembly history is characterised by the epoch $a(m)$ at which a fraction $m$ of the present-day mass has been first assembled.
    The \emph{absolute values} of the otherwise negative correlation coefficients are shown in this panel, and $m$ is shown in logarithmic scale.
    In the middle panel the mass assembly history is alternatively characterised by the fraction of the present-day mass, $m(a)=M(a)/M(a=1)$, that has been assembled by the time of each $a$, and in the bottom panel by $\Delta m(a)$, the step-wise increment in $m$ at each $a$.
    The top $x$-axes of the middle and bottom panels show the corresponding redshift and age of the universe respectively.
    In all three panels, the different colours represent results for the different mass samples, as is labelled in the top panel.
    The solid lines show the Spearman's rank-order correlation coefficients between $\Mark(c)$ and $\Mark(a(m))$ (top panel), $\Mark(c)$ and $\Mark(m(a))$ (middle panel), or $\Mark(\Delta m(a))$ (bottom panel).
    Each horizontal dashed line in the top panel shows the Spearman's correlation between $\Mark(c)$ and the mark value of the optimal linear combination of $a(m)$'s for the corresponding mass sample, while the horizontal dashed lines in the middle panel indicate the Spearman's correlations between $\Mark(c)$ and the mark values of the optimal linear combinations of $m(a)$'s. These optimal linear combinations contain most but not all of the information about the present-day concentration.
    }
    \label{fig:spearmanr}
\end{figure}

\subsection{Correlation with Mass at a Specific Time}
\label{sec:stepwise}

There have been several attempts to summarise the mass assembly history 
with a single parameter that correlates strongly with the 
present-day concentration \citep[e.g.,][]{bullock01,wechsler02,zhao03,Correa2015II}.
Two common choices are the halo 
half-mass scale, $a_{1/2}$, which is the epoch at which the halo 
first assembled half of its present-day mass, and the W02 
formation time, $a_c$, which serves as an estimate of the end of the 
early phase of rapid mass accretion by the halo. 
These attempts were relatively successful, suggesting that 
much of what determines contemporary halo concentration can 
be summarised with one quantity and that there may exist a 
``key stage'' in a halo's assembly history that 
substantially impacts the halo's internal structure.

This motivates us to conduct a systematic, empirical 
search for the stage of mass assembly that is most 
correlated with the present-day concentration $c$. 
We quantify mass assembly histories in two ways: 
(1) by the epoch $a(m)$ at which a fraction $m$ of the 
present-day halo mass is first assembled 
(for example, the half-mass scale $a_{1/2}=a(m=0.5)$); and 
(2) by the relative mass fraction $m(a) = \Mvir(a)/\Mvir(a=1)$, 
which is the mass of the halo at time $a$ 
in units of its contemporary mass. 

For the purposes of this paper we choose to represent the mass assembly histories using the mass of main progenitors as a function of scale factors.
However, there are multiple other characterisations of the formation history that we have not explored, for example, the collapsed mass history \citep[e.g.,][]{nfw97,gao08,ludlow2012}, and the transition between rapid and slow accretion phases \citep[e.g.,][]{zhao03}.

Both concentration and mass assembly history are 
known to correlate with present-day mass. 
While we work with mass-selected halo samples, 
we further mitigate any correlations 
induced by the mass dependence of the relative mass fraction and 
concentration as follows \citep[see][]{mao_etal18}. 
We divide each mass-selected halo sample 
shown in \autoref{tab:mass_sample} 
into narrow bins. Within each of these bins, 
we assign each halo a mark, $\Mark(x)$, where 
$x$ is the property of interest. Either 
$x=c$ or $x=m(a)$ in our present discussion. 
$\Mark(x)$ is the percentile rank among all  
of $x$ within the bin. For example, $\Mark(x)$ 
ranges between $\Mark(x)=0$, for the halo with 
the lowest value of $x$ in the bin, and 
$\Mark(x)=1$, for the halo with the highest 
value of $x$ in the bin. Each of our three 
mass-selected samples corresponds to a range of 
halo masses given in \autoref{tab:mass_sample}. 
We divide the $10^{12}\, \Msunh$ and $10^{13}\, \Msunh$ 
samples into 20 logarithmically-spaced mass bins and the 
$10^{14}\, \Msunh$ sample into 30 logarithmically-spaced bins.

We study correlations with the concentration mark $\Mark(c)$, for the two forms of mass assembly history, $\Mark(a(m))$ and
$\Mark(m(a))$, as defined in the 
preceding paragraph. Specifically, we compute $\Mark(m(a))$ for the 99 values of $a$ that correspond to the 99 available 
snapshots of the simulations, and $\Mark(a(m))$ for $m=\left \{0.01,0.02,0.03,...,0.99, 1\right \}$.
We then calculate
the Spearman rank-order correlation, $\rho$, 
between these marks of the assembly history and $\Mark(c)$.

The Spearman rank-order correlations between 
$\Mark(a(m))$ and $\Mark(c)$ as a function of 
the fraction $m$ are shown in the top panel of \autoref{fig:spearmanr}. The lines of different colours 
represent the different mass samples, 
as labelled in the same panel. 
The correlation coefficients are negative for all the values of $m$, and we show them in absolute values.
This is in accordance with previous understanding that haloes 
are likely to be more concentrated if they assembled their 
masses at smaller scale factors.
For all three samples, the correlations at large mass fractions are smaller than those at both medium and small fractions.
The $a(m)$'s defined at a range of medium mass fractions ($0.3\lesssim m \lesssim 0.7$) contain similar and relatively high levels of information about the present-day concentration.
This also explains the comparable effectiveness of various definitions of formation time in previous literature.
On the other hand, it is obvious from the figure that the time at which the main progenitor of a halo gains a low fraction of its final mass \citep[e.g., 4\% as in][]{zhao_etal09} is not as informative as medium mass fractions, such as the commonly used half-mass scale, $a_{1/2}$.

The Spearman correlations between $\Mark(m(a))$ and $\Mark(c)$ as a function of $a$ are shown as solid lines in the middle panel of \autoref{fig:spearmanr}.
The positive correlation at all times before $a=1$ 
is also consistent with earlier-forming haloes being more concentrated. The correlations for all three samples are 
relatively low at early and late epochs, and peak between 
$a \approx 0.3$ and $a \approx 0.7$, depending upon halo mass. 
By construction, $m(a=1)=1$ in all cases, so all 
correlations converge to 0 at $a=1$.
The peak of the correlation curve, 
which indicates the epoch at which the relative mass fraction $m(a)$ 
is best correlated with concentration, occurs later for more massive 
haloes. This is consistent with the tendency of more massive haloes 
to form later, so that if there is an important epoch in the 
evolution of a halo that influences its internal structure, it too 
occurs later for more massive haloes. 

The significant correlation 
between the concentration and the two characterisations of mass assembly history
is in broad accordance with previous studies that identify 
formation epochs of haloes that influence halo concentration.
However, notice that the correlation curves in both the top and the middle panels
peak at $\rho \lesssim 0.7$, suggesting that
factors in addition to the mass of a halo at a particular 
time contribute to contemporary halo concentration. 
We will investigate this further below.

\subsection{Correlation with Mass Change at a Specific Time}
\label{sec:stepwise_dm}

The values of mass fraction, $m(a)$, at successive time steps are 
strongly correlated with each other, and the resulting 
correlation coefficients in the top panel of \autoref{fig:spearmanr} are 
not independent. To resolve the relative importance of 
\emph{instantaneous} mass growth at different epochs, 
we repeat the analysis in \autoref{sec:stepwise} 
for the increment in mass fraction between adjacent snapshots, 
$\Delta m(a)$, instead of $m(a)$, with $\Delta m(a_i) = m(a_{i+1})-m(a_i)$.

The correlations between instantaneous mass acquisition, 
$\Mark(\Delta m(a))$, and concentration, $\Mark(c)$, 
are shown in the bottom panel of \autoref{fig:spearmanr}. 
It is evident that earlier growth is positively correlated with concentration (with the exception of 
the very earliest snapshots at which time the haloes are 
poorly resolved), 
while later growth exhibits anti-correlation. 
Similar to \autoref{sec:stepwise}, earlier growth is less 
informative for more massive haloes. Moreover, 
$\rho$ peaks at lower values for more massive haloes indicating 
that their early assembly histories generally have less information 
on concentration compared to haloes of lower mass.

The peaks of the correlation curves in all panels of 
\autoref{fig:spearmanr} are broad. 
This indicates that a wide variety of times during 
the formation of a halo provide similar amounts of information 
on contemporary halo concentration. This is likely why a 
variety of halo formation time measures, such as $a_{1/2}$ and 
$a_c$, show similar levels of correlation with present-day 
halo concentration. The breadth of the peaks in 
\autoref{fig:spearmanr} further suggests that one cannot choose a 
single definition of the formation time that will dramatically 
outperform a variety of other reasonable choices. The 
distillation of the mass assembly history into a single parameter 
inevitably leads to a significant loss of information.

At late times, the correlation between concentration and mass increase 
becomes negative and reaches a minimum at $a \approx 0.83$ for all 
three mass samples. This is suggestive that the same dynamical 
process has caused this behaviour.
In \autoref{sec:pseudo-phys} below, we identify this dynamical 
process to be mergers. Mergers account for the anticorrelation 
in general, the stronger anticorrelation between $\Mark(c)$ and 
$\Mark(\Delta m(a))$ for more massive haloes, and the uneven 
feature at $a\approx0.9$.

\subsection{Linear Regression of Mass Assembly History to Predict Concentration}
\label{sec:lin_comb}

Efforts to explain concentration with mass assembly history are often focused on singling out a formation time that best represents the mass assembly history.
However, we have shown in the previous subsections that multiple epochs in the mass assembly history contain similar amounts of information on concentration, which disfavours a single definition of formation time for this purpose.
In order to integrate information on concentration from the full mass assembly histories, we perform an ordinary least squares linear regression with $\textit{\textbf{a}}=\left \{ a(m=0.01), a(m=0.02), ..., a(m=0.99), a(m=1)\right \}$ as the predictor variables and $\Mark(c)$ as the outcome variable, by fitting for a set of linear coefficients that minimises $\sum\limits_{n}\left (A_0+\sum\limits_{i}A_i a_i - \Mark(c)\right )^2$, where $\sum\limits_n$ is the sum over all haloes,  $\sum\limits_i$ is the sum over all values of mass fraction $m$ at which $a(m)$'s are defined, 
and $A_0$ and $A_i$ are the linear coefficients.
Similarly, we fit a set of linear coefficients, $B_0$ and $B_i$, with $\textit{\textbf{m}}=\left \{ m(a=0.06),m(a=0.065),...,m(a=0.99),m(a=1)\right \}$ as the predictor variables, that minimises $\sum\limits_{n}\left (B_0+\sum\limits_{i}B_i m_i - \Mark(c)\right )^2$, but here $\sum\limits_i$ is the sum over all snapshots 
(i.e. over all values of $a$).
For the present study, we elect to perform a simple linear regression, 
and refrain from more sophisticated forms of regression, 
because mass assembly histories of individual haloes are both 
volatile and noisy and these properties introduce the possibility 
of unphysical overfitting. For this reason, more complex 
regression methods warrant further, dedicated study.

We compare the results of the linear regression to the 
results of the previous section as follows. 
We determine 
the set of coefficients, $A_0$ and $A_i$, that gives the linear combination 
of the elements of $\textit{\textbf{a}}$ that is the most 
strongly correlated with $\Mark(c)$.
We repeat the process for the elements of $\textit{\textbf{m}}$ to obtain the optimal coefficients $B_0$ and $B_i$.
We then calculate the Spearman's correlations between 
$\Mark(c)$ and the marks corresponding to the resulting
linear combinations.

The correlation coefficients for the optimal linear combinations of the two characterisations of mass assembly histories are shown 
in the top and middle panel of \autoref{fig:spearmanr} respectively, as horizontal dashed lines. 
For both the set of $a(m)$'s and $m(a)$'s, and at all three masses, even the optimal linear combinations 
leave much of the dependence of concentration on mass assembly history unexplained, though they exhibit moderate improvements upon the best performing single parameters.
Performing the linear regression with 
$\log\textit{\textbf{m}}$ instead of $\textit{\textbf{m}}$ yields 
similar, but slightly weaker correlations.

For comparison, we measure the mark correlation between the concentration and the formation time defined as the epoch at which the mass of the main progenitor first reaches the characteristic mass enclosed within the scale radius at the present-day, $M_s=M(r<r_s)$.
We find that the level of correlation for this formation time is similar to the optimal linear combination of $m(a)$'s for all three of our mass samples.
We further note that this definition of the formation time is not independent of the concentration measured at the present-day, as it requires knowledge of $r_s$, and therefore is different from the proxies of the assembly history that we have employed so far.
This comparison suggests that our optimal linear combination captures most of the information in different forms of the formation history on the main branch.

In \autoref{sec:pseudo-phys}, 
we explore the combined effect of merger events happening at 
different times on halo structure.
We show that this combined effect cannot be described linearly.

\section{Concentration from Pseudo-evolution and Mergers}
\label{sec:pseudo-phys}

In the previous section, we attempted 
to explain contemporary halo concentrations 
using halo mass assembly histories. 
The incomplete success of this endeavour 
prompts further inquiry into additional factors in the evolution 
of haloes that may influence halo density profiles. 
We expect the density profile of a halo to be largely determined by 
the halo's prior mass assembly history, 
independent of the redshift 
at which the halo is observed.
We therefore extend our investigation 
to the study of the full evolution of halo concentrations, 
and search for connections between 
the behaviour of halo concentrations 
and events in halo mass assembly histories.

In this section, we study the evolution of halo concentration $c$, 
and halo scale radius $\rs$, both during quiescent periods of halo 
pseudo-evolution and during merger events. 
We find that halo structure undergoes significant changes 
in response to major, and even minor, mergers in a manner 
that is qualitatively universal. We propose a physical 
explanation for the response features that we observe. 
We further propose that the scale radii and concentrations 
of haloes result from pseudo-evolution punctuated by 
marked fluctuations associated with merger activity.

\begin{figure*}
    \centering
    \includegraphics[trim=70 50 0 80, clip, scale=0.44]{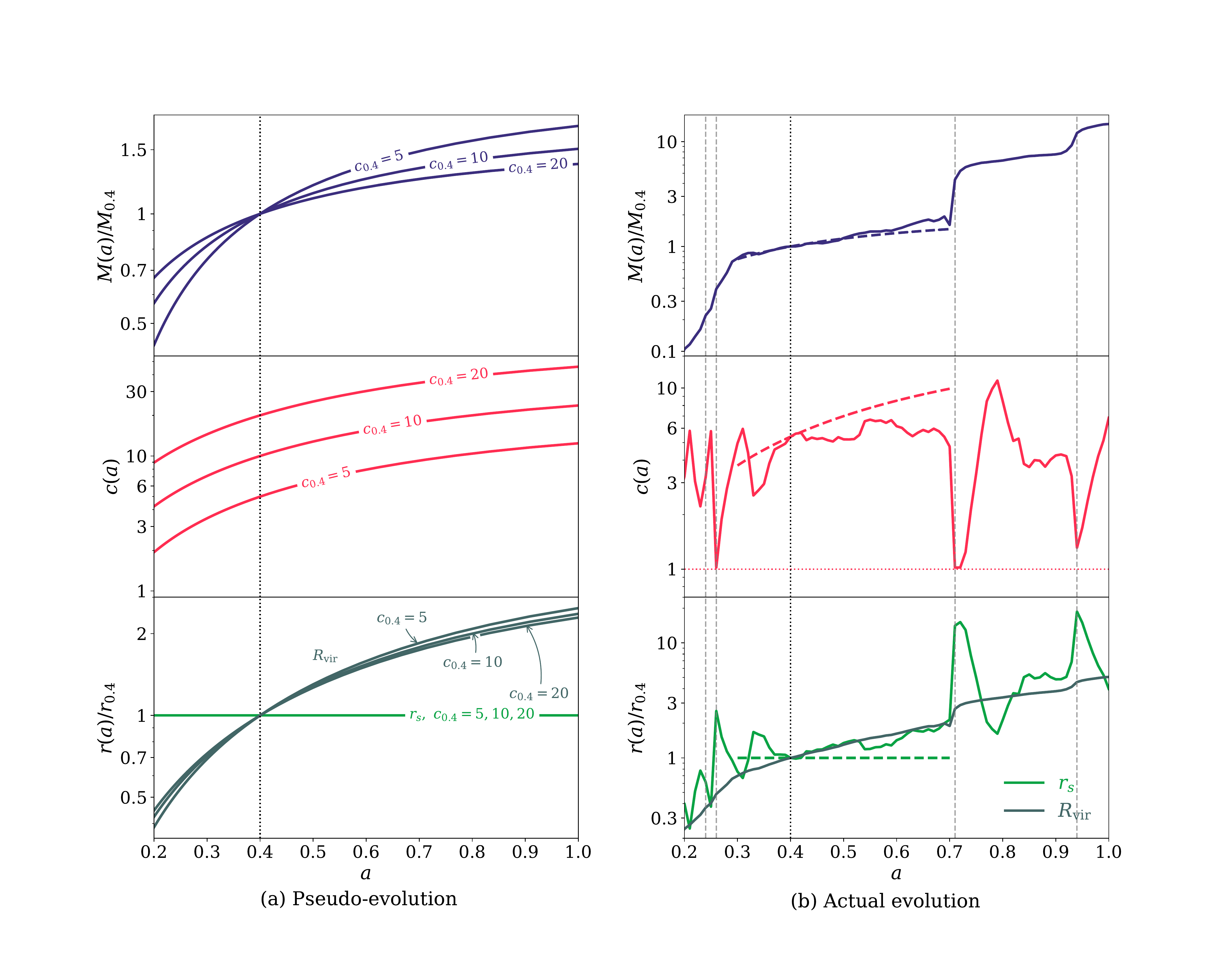}
    \caption{
    (a) In the left column, we show the change in mass (top panel), concentration (middle panel), and virial radius and scale radius (bottom panel) due to pseudo-evolution, choosing $a=0.4$ as the reference state, denoted with the subscript ``0.4'', and marked by vertical black dotted lines in the panels.
    In the bottom panel, the $y$-axis indicates $r(a)/r_{0.4}$, where $r$ is either the virial radius $\Rvir$ or the scale radius $\rs$.
    The change in mass and radii are plotted in terms of the ratio between the pseudo-evolved values and the values at $a=0.4$.
    Each line of pseudo-evolution is labelled with the corresponding concentration at the reference point $a=0.4$, as the ratios and concentration are only functions of the concentration, independent of halo mass.
    It can be observed that significant growth in both halo mass and halo concentration can be associated with pseudo-evolution.
    (b) The right column is similar to the left column, but shows the actual evolution of an individual halo's mass, concentration, and virial radius and scale radius, as functions of the scale factor $a$.
    The vertical gray dashed lines mark the major mergers identified by \textsc{Consistent Trees}, and the horizontal dotted line in the middle panel indicates $c=1$ to guide the eye.
    Besides the actual evolution, we also show the pseudo-evolution for comparison.
    In each panel, the dashed curve of the same colour as the solid curve shows the corresponding quantity pseudo-evolved from the state at $a=0.4$ (vertical black dotted line), between $a=0.3$ and $a=0.7$.
    In the bottom panel, only the pseudo-evolution of $\rs$, which is a constant function of time, is shown, while the pseudo-evolution of $\Rvir$ is omitted for clarity.}
    \label{fig:M-c-R}
\end{figure*}

\subsection{The Pseudo-evolution of Halo Mass and Concentration}

Pseudo-evolution refers to the fact that 
halo masses, virial radii, and concentrations all evolve 
even in the absence of merger activity or changes to scale radii 
\citep{diemer_etal13}. 
This is because haloes are traditionally defined to be regions 
with a mean density larger than $\sim$50--100 times the critical 
density (or $\sim$200--350 times the background density). 
As cosmological expansion dilutes the universe, the size of the 
region above the density threshold increases even in the absence 
of any coherent, inward flow of mass. Consequently, 
$c = \Rvir/\rs$ grows because $\rs$ remains approximately 
constant in the absence of significant merger activity.

In the left column of \autoref{fig:M-c-R}, 
we show the pseudo-evolution 
of halo mass, concentration, virial radius, and scale radius 
between $a=0.2$ and $a=1$, calculated using the \textsc{Colossus} 
software package \citep{colossus2018}, and assuming NFW profiles.
Each panel depicts halo properties evolved both forward and backward 
from an initial point of $a=0.4$ assuming pure pseudo-evolution. 
The pseudo-evolution is, itself, a function of halo concentration 
and we show halo pseudo-evolution for three different initial 
concentration values, $c_{0.4}\equiv c(a=0.4)=5,10,20$, 
in each panel. 
The \emph{top panel} shows the pseudo-evolved 
mass normalised by the mass 
at $a=0.4$, $M(a)/M_{0.4}$, 
which is only a function of the concentration, 
independent of halo mass.
In the \emph{middle panel}, 
we show the pseudo-evolution of concentration for the three values 
of $c_{0.4}$ separately. The evolution of 
$c(a)$ under pseudo-evolution is also independent of mass. 
The bottom panel shows the evolution of the scale 
radius $\rs(a)/r_{\mathrm{s},0.4}$ and virial radius 
$\Rvir(a)/R_{\mathrm{vir},0.4}$ in physical units 
where $r_{\mathrm{s},0.4}$ is the scale radius evaluated 
at $a=0.4$ and likewise for 
$R_{\mathrm{vir},0.4}$.  
These ratios are also independent of mass, 
and since the physical $\rs$ remains constant under 
pseudo-evolution, the ratio 
$\rs(a)/r_{\mathrm{s},0.4}=1$ independent of $a$. 
In all of the panels, the lines are labelled by 
the corresponding $c_{0.4}$ values.

The left panels of 
\autoref{fig:M-c-R} show that pseudo-evolution contributes 
substantially to the evolution of halo size and concentration in 
the absence of any physical mass inflow or accretion. 
In the following subsections, 
we study the effect of physical accretion, 
which includes all the merger activities 
beyond pseudo-evolution.


\subsection{Case Study: The Co-evolution of Halo Mass, 
Concentration, and Scale Radius During Mergers}
\label{sec:m_and_c_evo}

In the case of pure pseudo-evolution, 
the evolution of halo mass, virial 
radius, and concentration from some initial state can be predicted. 
Significant deviations from the predictions of pseudo-evolution can 
likely be attributed to the physical inflow of mass across the 
virial boundary of the halo. 
To investigate how deviations from pseudo-evolution 
affect halo structure, we begin with a case study.

In the right column of \autoref{fig:M-c-R}, 
we show the evolution of mass, 
concentration, virial radius, and scale radius for an 
individual halo. 
We neglect the evolution before $a=0.2$, 
which is relatively poorly resolved.
For consistency, we use the same quantities 
as in the left column, i.e., 
the concentration, as well as the mass and radii 
normalised by the values at $a=0.4$, 
but note that the ranges of the $y$-axes are different.
In the middle panel, $c=1$, the lower boundary of fitted halo concentration, is marked by the horizontal dashed line.
Major mergers in the mass assembly history of this halo 
are marked by gray, vertical, dashed lines.

Notice in the top panel that this particular 
halo undergoes no major mergers between $a \approx 0.25$ and 
$a \approx 0.7$. During this relatively quiescent period in the 
halo's mass accretion history, the halo's mass evolution is quite close 
to that predicted by pseudo-evolution, which we show with a dashed line for comparison. As in the left column of \autoref{fig:M-c-R}, the pseudo-evolution is 
computed from $a=0.4$. 
In the middle and bottom panels, 
the pseudo-evolution of the concentration and the scale radius during this period 
are also shown as dashed lines. 
The evolution of both the concentration and the scale radius during the period between 
major mergers is relatively mild. Comparing the actual evolution 
of these properties to the predictions of pseudo-evolution 
reveals non-negligible differences. Furthermore, decreases 
in the actual evolution of halo concentration seem to be visually 
associated with small deviations in the mass assembly history 
that are not identified as major mergers.
This suggests that even small amounts of physical
mass accretion can lead to significant deviations
from the pseudo-evolution of concentration and scale radius. 
This, in turn, suggests that the scatter in the 
profiles of a population of haloes may be caused 
by small differences in mass assembly histories.

Focus now on the major merger events in \autoref{fig:M-c-R}. 
Prominent features can be observed in the temporal neighbourhood
of each major merger event. Concentration decreases rapidly 
and significantly to a minimum at approximately the time of 
the major merger. Subsequent to the merger, concentration 
immediately increases, decreases again, and then stabilises. 
After stabilising, there is a long period of secular increase 
of halo concentration.
The change in concentrations due to major mergers is large 
compared with 
the scale of the overall concentration evolution throughout 
the entire history of the halo. 
Meanwhile, the scale radius follows the same trend but 
in the opposite sense, as is expected.

In the bottom panel, it is obvious that the change in 
$\Rvir$ is much less dramatic and much simpler than that 
in $\rs$ in response to major mergers.
$\Rvir$ increases due to both pseudo-evolution and 
the physical increase in mass, 
while $\rs$ remains constant unless the inner profile is impacted.
Concentration is the ratio $c=\Rvir/\rs$. As is now apparent, 
discussing this ratio complicates our discussion unnecessarily, 
because the two radii have very different mechanisms of evolution. 
$\Rvir$ evolves rather modestly and in approximate correspondence 
with predictions from pseudo-evolution along with mass increases 
due to mergers. Large changes in concentration are induced by 
the large deviations in $\rs$ brought about by mergers. 
We will therefore focus on the scale radius $\rs$ 
instead of the concentration for the rest of 
this subsection.

\begin{figure*}
    \centering
    \includegraphics[trim=20 10 10 10, clip, scale=0.55]{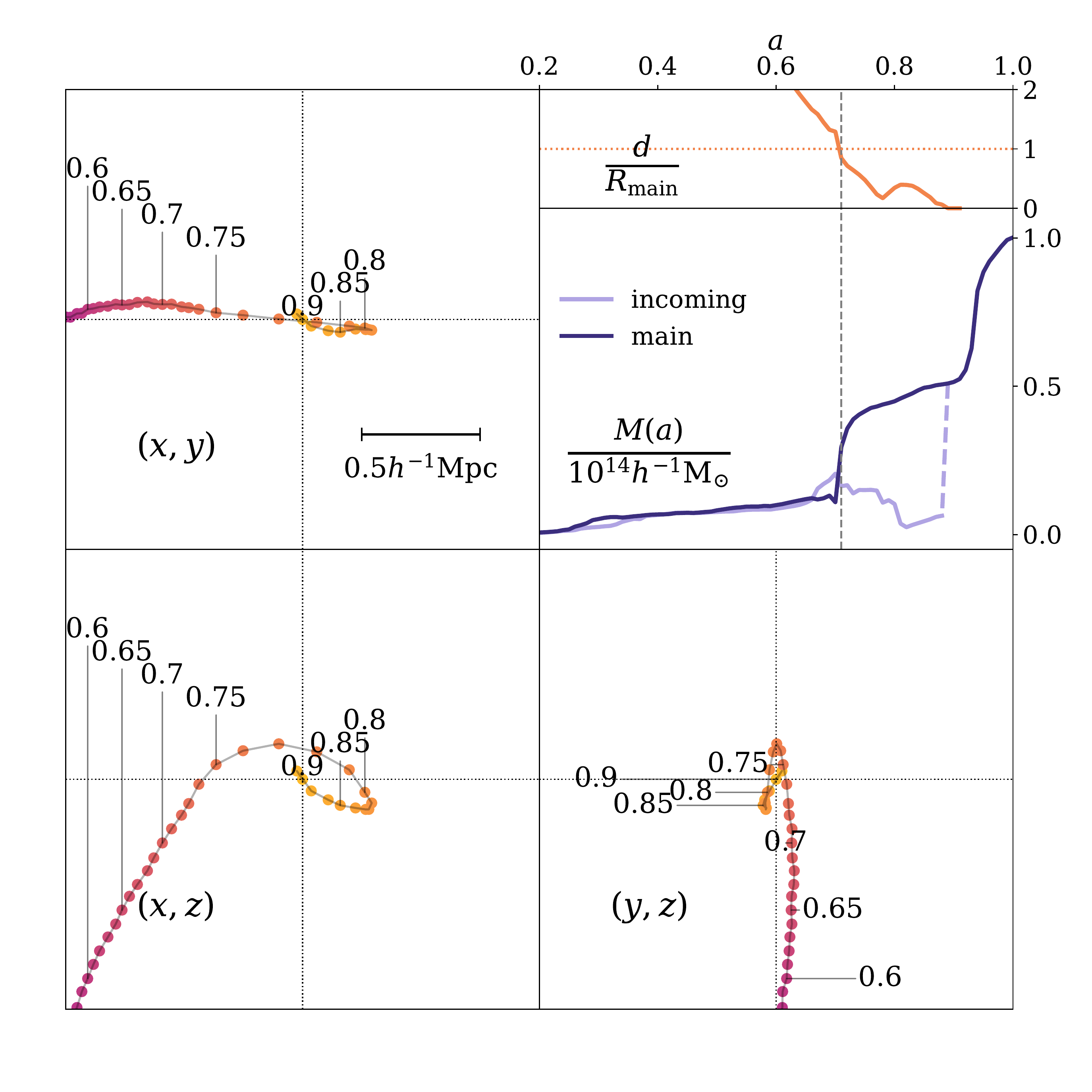}
    \caption{In this figure, we show the process of the major merger that the halo in \autoref{fig:M-c-R} undergoes at $a=0.71$.
    The top left and bottom panels show the orbit of the incoming progenitor around the main progenitor in the three projected planes respectively, displaying the comoving space from $-1\Mpch$ to $1\Mpch$ in each direction, and the comoving length scale of $0.5\Mpch$ is shown in the upper left panel for visual clarity.
    Each point in an orbit represents the state in a different snapshot, colour coded from dark to bright with the increase of time; the scale factor $a$ is labelled at several points.
    The upper right panel shows the time span between $a=0.2$ and $a=1$, with the time of the major merger marked by a gray vertical dashed line.
    The lower part of this panel tracks the mass changes of the main and incoming progenitors in units of $10^{14}\Msunh$.
    The incoming halo's evolution ends when it is completely disrupted and is no longer identified as an object, and the transition is shown as a dashed line. 
    The increase in mass in the main branch afterwards is due to another major merger that follows.
    The upper part of the same panel shows the evolution of $d/R_{\rm{main}}$, where $d$ is the distance between the centres of the two progenitors, and $R_{\rm{main}}$ is the virial radius of the main progenitor.
    The ratio $d/R_{\rm{main}}$ decreases below 1 at around $a_{\rm MM}$, marked by the horizontal dotted line, and reaches 0 as the incoming object disappears.}
    \label{fig:orbit}
\end{figure*}

We take the major merger at $a=0.71$ as an example 
to discuss the common features, 
and interpret the response in $\rs$ with the 
dynamical processes that occur during the major merger event.
The two progenitors of this major merger are examined in 
\autoref{fig:orbit}, 
which illustrates the orbit of the incoming progenitor 
around the main progenitor, 
as well as the mass evolution of the incoming and 
main progenitor haloes.
During the major merger, the incoming progenitor 
loses mass to the main progenitor before being 
completely disrupted.
Without significant physical mass growth, 
the scale radius only varies slowly, 
which can be seen in the period prior to the major merger 
at $a=0.71$ in \autoref{fig:M-c-R}, 
where the mass growth of the halo is mainly due 
to pseudo-evolution.

Notable deviations from the pseudo-evolution of the 
halo scale radius can be seen as the incoming halo 
traverses the main progenitor halo. 
As the merger begins, the halo's scale radius departs from 
its original evolution, and quickly increases, 
approaching the physical boundary $\rs=\Rvir$, 
which suggests essential deviation from an NFW profile.
This is due to the incoming progenitor entering the 
virial boundary of the main progenitor, 
shown in the upper part of the upper right panel in 
\autoref{fig:orbit}, 
placing a relatively large amount of mass in the 
periphery of the main halo and rendering 
the outer profile shallower. 
The shallower spherically-averaged density profile yields a 
larger scale radius.
Later, as the incoming halo approaches the centre of the main 
halo, the scale radius falls because mass is then inordinately 
concentrated near the halo centre. The scale radius increases 
again as the merging halo moves outward from the centre of 
the main halo on its orbit. Compared to the secular evolution 
in scale radius seen during quiescent periods, this evolution 
of halo scale radius is rapid, occurring over approximately 
one halo crossing time.

The merger concludes with the incoming halo spiralling inward toward 
the centre of the main halo due to dynamical friction. As this 
happens, the scale radius once again increases. 
The incoming object is gradually disrupted, 
and $\rs$ resumes secular evolution.
The recovery after the major merger at $a=0.71$ 
is interrupted by a later major merger that follows at 
$a=0.94$; however, the recovery process can be observed 
in \autoref{fig:M-c-R} after the major merger 
at $a=0.26$.

With this example, we have shown that during a major merger event, 
$\rs$ experiences consequential changes, that can be attributed 
to the dynamical processes of the progenitors.
Our interpretation is in agreement with \citet{ludlow2012}, who also observed the oscillations in a halo and related them with the crossings of the merging object before virialisation.
These changes are extended in time, motivating an investigation 
of the time scales that are involved in the next section.

\subsection{Universality of Response}
\label{sec:rs_uni}

In the previous subsection, we followed the co-evolution of mass, 
concentration, and scale radius of one halo, 
focusing on the dynamical processes associated 
with major mergers that drive the evolution of halo scale radius. 
Based on this case study, we argued 
that halo scale radii respond to mergers in an 
oscillatory manner and that the oscillations are due 
to orbital evolution. Accordingly, 
it is natural to study the evolution of haloes due 
to mergers with time measured in units of the 
local dynamical time, 
the time required to orbit across an equilibrium 
dynamical system, in our case a halo.
We adopt the definition of the dynamical time
\begin{equation}
    \tdyn=\sqrt{\frac{3\pi}{16G\bar{\rho}}},
    \label{eq:tdyn}
\end{equation}
where $G$ is the gravitational constant and $\bar{\rho}$ 
is the mean density of the system, which we choose to be the virial density of haloes.
With a given cosmology, 
the dynamical time $\tdyn$ is dependent on the 
scale factor $a$ through $\bar{\rho}$.
In the cosmology adopted by the Dark Sky Simulations, the 
dynamical time scales as 
\begin{equation}
\tdyn \approx 3.15 (1+z)^{-3/2}\, \mathrm{Gyr}.
\end{equation}

Following \citet{Jiang_vdB16}, we then define a new quantity $T$, 
which measures the time between two epochs in units of the dynamical 
time, as
\begin{equation}
    T(a;a_{\rm{ref}}) = \int_{t(a_{\rm{ref}})}^{t(a)}\frac{dt}{\tdyn(t)},
    \label{eq:Ntau}
\end{equation}
where $t(a)$ is the age of the Universe corresponding to the 
scale factor $a$, and $a_{\rm{ref}}$ is the reference epoch. 

To study the general behaviour of major mergers, we select haloes from the simulation that undergo major mergers along the main branch, independently of their masses.
The major merger times we select are 
$a_{\rm MM}=0.33,0.50,0.67,0.80$, 
corresponding to $z=2,1,0.5$, and 0.25 (see \autoref{tab:amm_sample}).

\begin{figure*}
    \centering
    \includegraphics[trim=20 0 50 40,clip,scale=0.5]{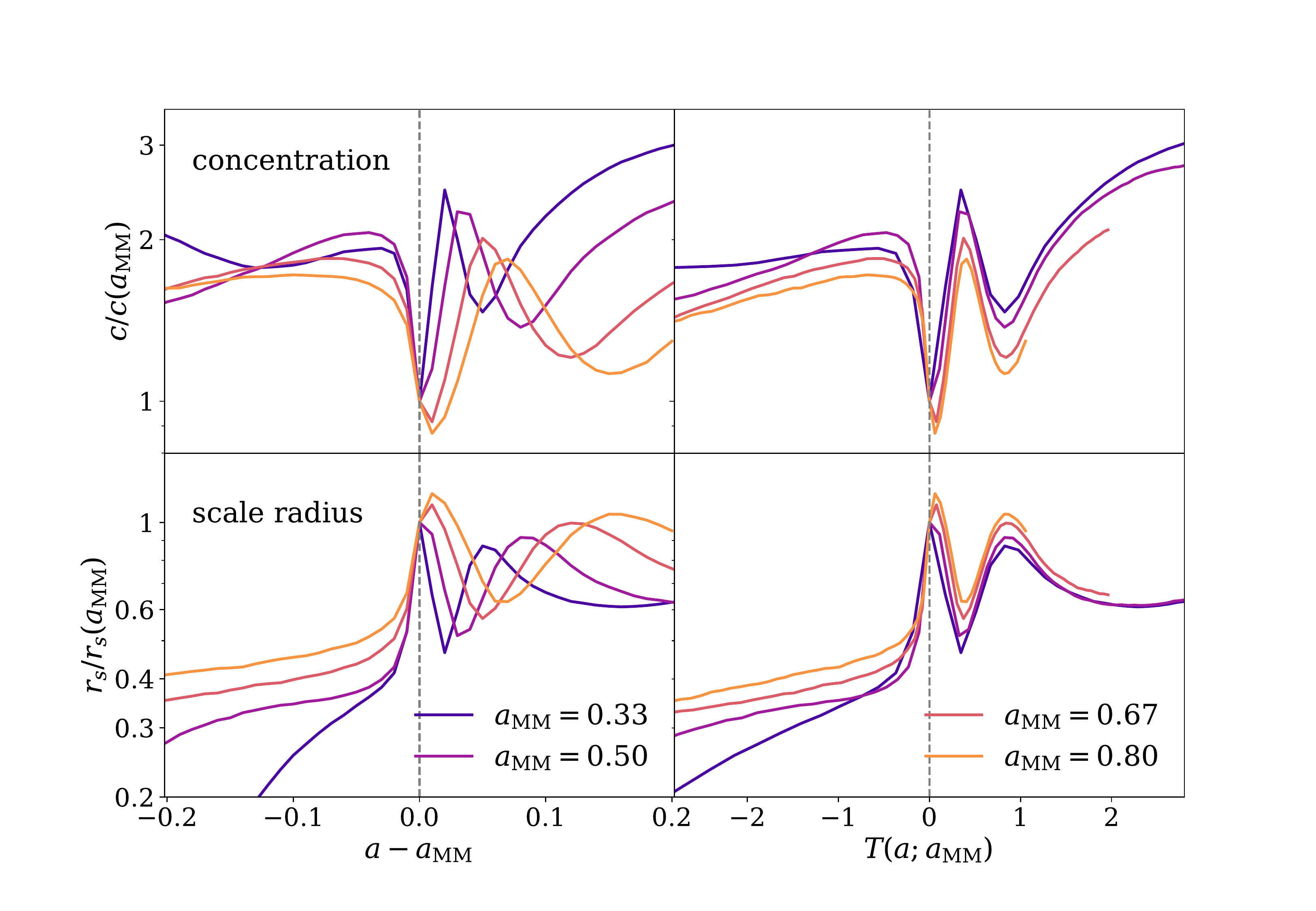}
    \caption{Median response to major mergers that happen at different times.
    The top row displays the concentration, and the bottom row displays the scale radius, both in logarithmic scale and normalised by the value at the time of merger.
    In the left column, time is measured in terms of the scale factor $a$, shifted with respect to $a_{\rm MM}$, while in the right column, time is measured in units of dynamical times, with the merger time as the reference point.
    The groups are colour coded by their respective $a_{\rm MM}$.
    The time of merger is marked by a vertical dashed line in each panel.
    In the right column, some of the lines are truncated due to the limited time range of the simulation.
    The response of haloes of different masses and major merger times are remarkably similar when scaled by dynamical time.}
    \label{fig:unitime}
\end{figure*}

We stack each $a_{\rm MM}$ group and examine the median evolution to reduce noise.
In \autoref{fig:unitime}, we show the median response of the concentration and scale radius in logarithmic scale, normalised by their respective values at $a_{\rm MM}$.
Time is measured both in terms of the scale factor 
and in terms of the number of dynamical times with respect to the time of merger.

In both columns, we observe the orbital features discussed in \autoref{sec:m_and_c_evo}, demonstrating that the dynamical processes shown in \autoref{fig:orbit} are universal, and that the incoming progenitor goes through one orbit on average before being disrupted \citep[see also][]{vdBosch17}.
However, only in the right column, where time is measured in units of dynamical times, are the responses from the different $a_{\rm MM}$ groups aligned, going through the oscillations with a remarkably universal dynamical timescale.
This further confirms the connection between the concentration and scale radius evolution and the dynamical processes during major mergers, as well as the universality of this mechanism when scaled with dynamical time.

\subsection{All Merger Activity}
\label{sec:all_mergers}

We have shown that the evolution of halo structure in response to major mergers have common features, with universal timescales measured in units of dynamical times.
The amplitude of the change in $\rs$ due to major mergers is large compared with the average scale of change over the entire history, and also much larger than that of the halo radius evolution, causing large fluctuations in halo concentration as well.
However, major mergers are relatively rare events.
The average numbers of major mergers between $a=0.25$ and $a=1$ for a halo are 1.14, 1.51 and 2.00 for the $10^{12}\Msunh$, $10^{13}\Msunh$ and $10^{14}\Msunh$ samples respectively.
Minor mergers with smaller ratios between progenitor masses happen much more frequently, and dominate the physical mass growth beyond pseudo-evolution.
As major mergers are the extreme cases of merger events, it is reasonable to expect that minor mergers have similar but less dramatic effects.

To examine the response to all merger activity, we search for instances of minor merger events in the random catalogue described in \autoref{sec:amm_sample}.
As the \textsc{Consistent Trees} code identifies major mergers only, we define minor mergers based on the rate of fractional mass increase between adjacent snapshots.
We calculate the rate of fractional mass increase as
\begin{equation}
    \Gamma(a_i) = \frac{\Delta M(a_i)/M(a_i)}{T(a_{i+1};a_i)},
    \label{eq:rate}
\end{equation}
where $\Delta M(a_i) = M(a_{i+1})-M(a_i)$ is the mass increase between the adjacent snapshots, and $T(a_{i+1};a_i)$ is the corresponding time interval in units of dynamical times. 
The rate of fractional mass increase, $\Gamma(a_i)$, is a dimensionless 
quantity.
The time interval $T(a_{i+1};a_i)$ for a fixed scale factor interval $a_{i+1}-a_i$ decreases as the Universe evolves, and drops below 0.2 by $a=0.33$, the first merger epoch we consider.
When selecting minor mergers, we consider the same epochs as for the major merger samples, $a=0.33,0.50,0.67,0.80$.
The mean values of $\Gamma(a_{\rm MM})$ for the major merger samples are 2.77, 2.77, 2.96 and 3.43 for the four major merger times respectively. 
At each of these time steps, we select our minor merger sample to have values of $\Gamma$ between $1.0$ and $1.5$.
We also require that there are no major mergers within $\pm0.25 \tdyn$ around the minor mergers\footnote{The mass increase associated with a major merger occurs over approximately $\pm 0.25\tdyn$ around the time of merger.}, to exclude mass increase associated with major mergers.

\begin{figure*}
    \centering
    \includegraphics[trim=20 0 50 35,clip,scale=0.5]{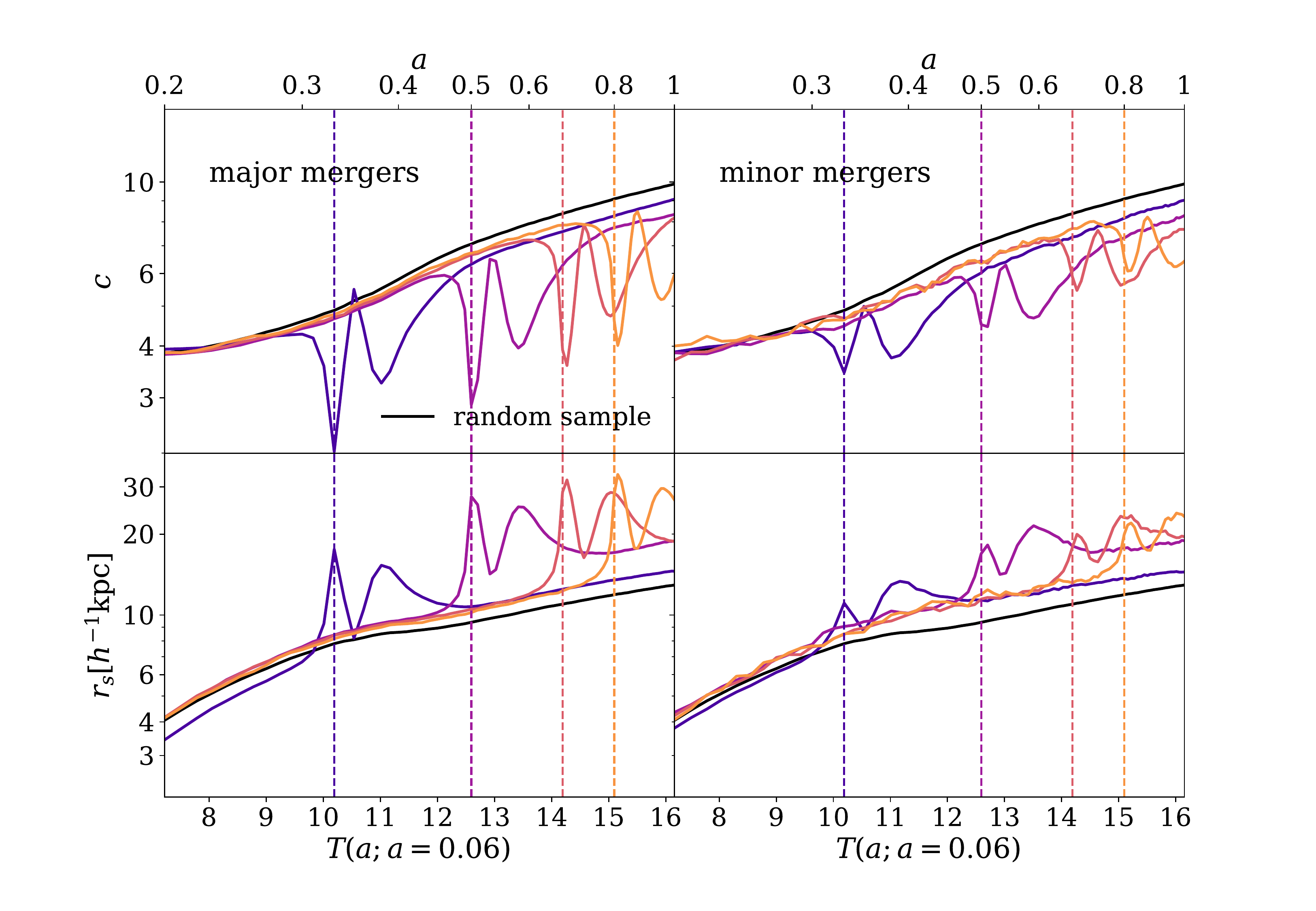}
    \caption{This figure compares major mergers with minor mergers that happen at the same epochs, which are marked by vertical dashed lines, each with the same colour as the corresponding evolution curve.
    Similar to \autoref{fig:unitime}, the halo evolution is tracked in terms of concentration in the top row and scale radius in the bottom row. 
    Time is measured in units of dynamical times, adopting $a_{\rm{ref}}=0.06$, and the corresponding scale factor $a$ is labelled at the top.
    The minor merger events are selected from the random catalogue by their rates of fractional mass increase $\Gamma$, defined in \autoref{eq:rate}, $1.0\leq\Gamma\leq1.5$, and no major mergers within $\pm 0.25\tdyn$ of the time step of interest.
    The left column shows the median evolution of each major merger sample, and the right column shows those of the minor merger samples.
    In every panel, the solid black curve depicts the median evolution of the random sample for comparison.}
    \label{fig:minor}
\end{figure*}

In \autoref{fig:minor}, we compare the haloes that undergo these minor mergers against the major merger samples and the randomly-selected halo sample.
The median evolution of each sample is plotted in terms of both concentration and scale radius, 
as functions of time.
In the bottom $x$-axes, time is represented as the number of dynamical times since the first available snapshot, while the corresponding scale factor is labelled on the top $x$-axes.
The major merger samples are shown in the left column, and the lines are colour coded according to the time of merger, marked by vertical dashed lines of the same colours.
The solid black curve shows the evolution of the random sample for comparison.
Similarly, the right column shows the minor mergers that happen at the same time steps.

It is obvious from \autoref{fig:minor} that minor mergers indeed cause qualitatively similar responses in the halo structure.
The magnitudes of these features, though smaller than those of the major merger response, are still significant compared with the scale of overall evolution throughout cosmic time.
This shows that all mergers, major or minor, involve similar dynamical processes, with the effect of expanding the inner profile and suppressing concentration during an extended period.
We also note that the haloes that undergo mergers have lower concentrations than the random sample of haloes even after several dynamical times, and we have tested that this difference in concentration cannot be accounted for by the difference in their mass distributions.
This could be due to the fact that mergers are correlated, perhaps due to environmental dependences, or that mergers have a persistent 
effect on the internal structures of haloes, or a 
combination thereof. 
Determining the nature of this effect is worthy of 
a distinct study in its own right.
The fluctuations in the scale radii and concentrations following minor mergers, which happen frequently for most haloes, are also a likely source of spread in the present-day values of halo internal properties, which we investigate in \autoref{sec:scatter}.

\subsection{Irreducible Scatter Due to Stochastic Mergers}
\label{sec:scatter}

With our improved understanding of mergers, 
we examine the role that these events play in 
producing the present-day concentrations and 
scale radii of halo samples with fixed masses.
We have shown in \autoref{fig:minor} that the impact on $\rs$ from major mergers and even minor mergers is significant compared with the scale of the overall $\rs$ evolution in the entire history, 
and persists over a considerable amount of time 
(several dynamical times, 
meaning several Gyr). 
Therefore, we expect that the cumulative response of 
a halo to the merger events in its 
mass assembly history is crucial to 
the determination of the scale radius and 
concentration of the halo. Moreover, both 
the relative sizes of mergers and the 
temporal distribution of these mergers 
are important in determining present-day 
concentration and scale radius.

\begin{figure}
    \centering
    \includegraphics[trim=0 40 20 80,clip,scale=0.43]{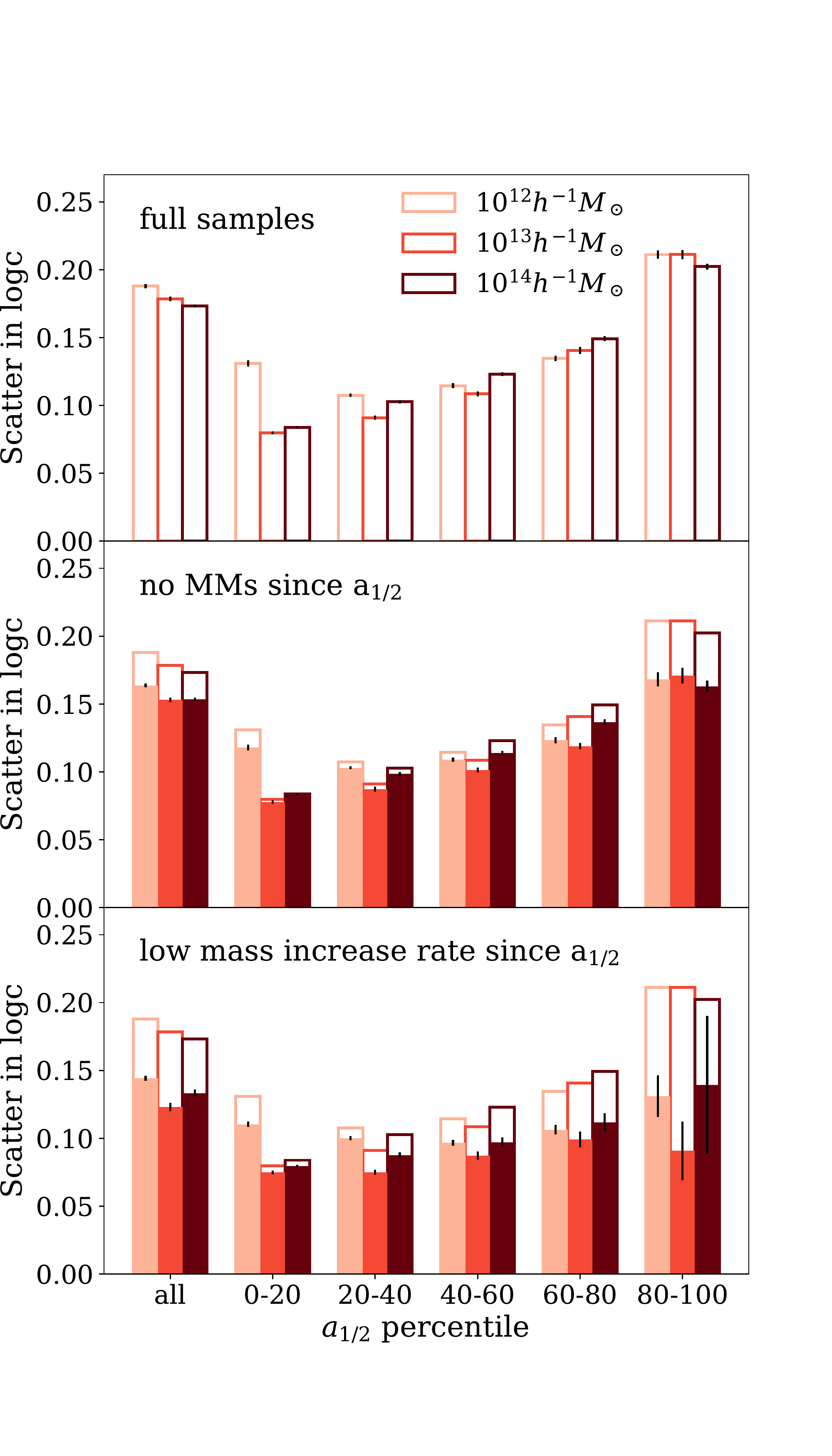}
    \caption{In this figure, we show the logarithmic scatter in concentration for the present-day mass samples.
    The present-day mass is colour coded and labelled in the bottom panel.
    In the top panel, the first group of bars shows the scatter for the entire samples, while the other groups are subsamples selected by their half-mass scale percentiles within each mass sample.
    The error bars are calculated using bootstrap resampling.
    The same bars are also shown in the two lower panels for visual guidance.
    The filled bars in the middle panel shows the scatter for the same samples, but excluding haloes that undergo major mergers after the half-mass scale.
    The filled bars in the bottom panel adopt a more stringent selection criterion, excluding haloes that have mass increase events with $\Gamma\geq1.0$.}
    \label{fig:sigmalogc}
\end{figure}

In the top panel of \autoref{fig:sigmalogc}, we show the scatter in $\log c$ for each present-day mass sample, and the remaining scatter after further dividing the samples into quintiles by $a_{1/2}$.
There is a scatter of approximately 0.1--0.2 dex in concentration with populations of haloes with both mass and $a_{1/2}$ fixed.
The scatter increases with later half-mass scales, as there are more recent merger events for these haloes.
This scatter originates from the variety of possible paths of mass assembly.
In the middle panel, we examine the same samples,  but exclude haloes that undergo major mergers since their half-mass scales.
The resulting scatter in $\log c$ decreases in every sample, which is consistent with our conclusion that major mergers contribute to the uncertainty in today's concentration.
The decrease is not as significant as one might naively expect from 
the large fluctuations in concentration caused by major mergers, 
because major mergers are rare events and 
impact a small fraction of the population.
In the bottom panel, we further exclude all haloes 
that have stepwise mass increases with $\Gamma\geq1.0$ 
since $a_{1/2}$, and the scatter is indeed further reduced.
That a more stringent restriction on mergers further reduces 
scatter in concentrations strongly suggests that it is the 
mergers themselves that drive a significant portion of the scatter. 
The dependence of the scatter on the half-mass scale is largely 
removed by excluding these mass increase events, 
which confirms that 
different frequencies of mergers are the cause of this dependence.
It is reasonable to expect that when even more 
stringent limits are put on the mass increase rate, 
the scatter will be further reduced; however,  
we are unable to test this explicitly due to limited sample sizes. As a supplement to \autoref{fig:sigmalogc}, in Appendix~\ref{sec:c-M_relation} we show the dependence of the concentration--mass relation on the half-mass scale.

\begin{figure*}
    \centering
    \includegraphics[trim=40 20 40 40,clip,scale=0.45]{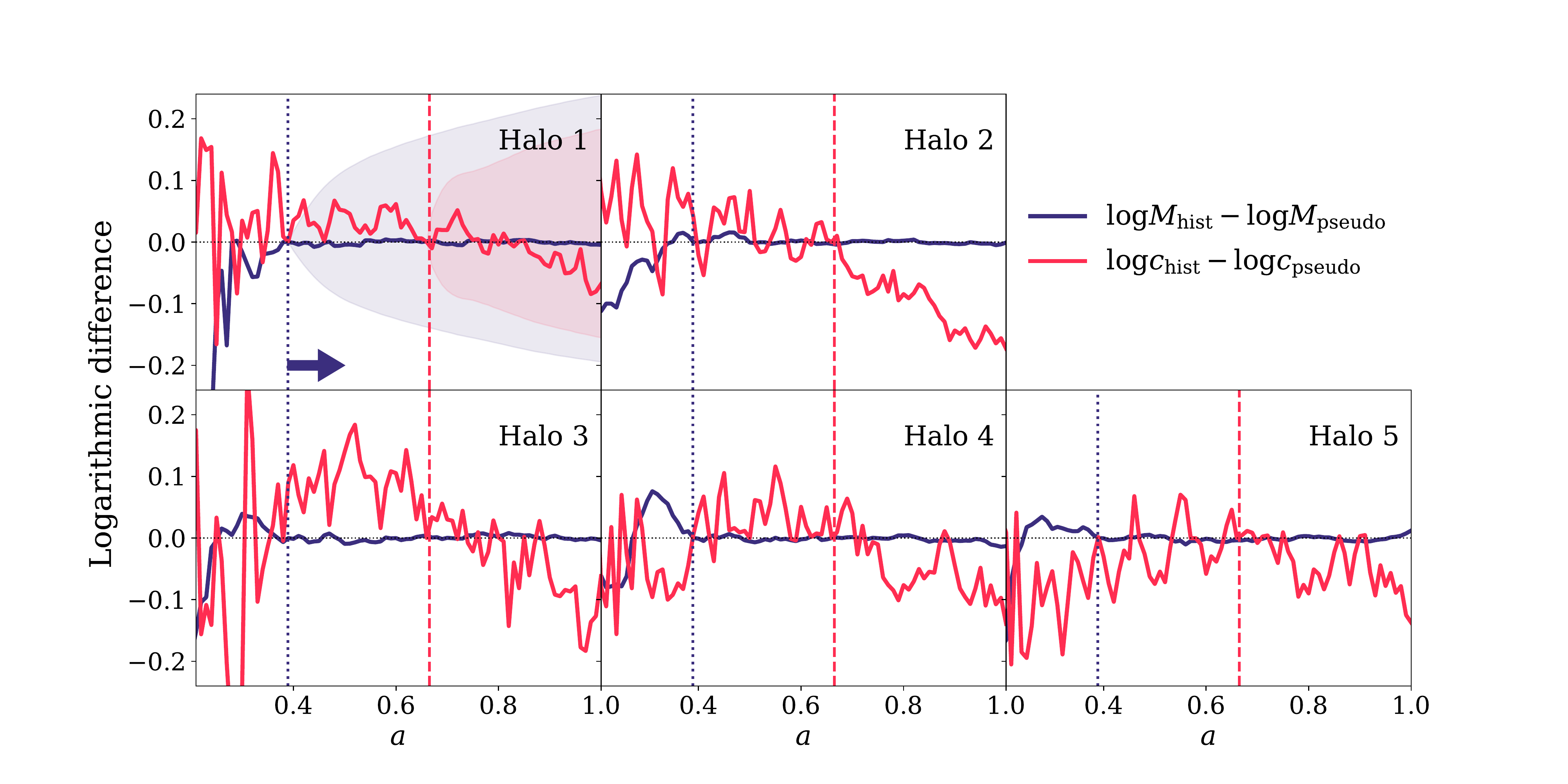}
    \caption{Comparison of the actual evolution against pseudo-evolution for five individual haloes in the simulation.
    These haloes are selected to have the least deviation from pseudo-evolution in mass in the last five dynamical times, which is marked by the vertical dotted dark blue line in each panel, and the dark blue arrow in the first panel.
    The concentration is compared against the forward pseudo-evolution from two dynamical times before $a=1$ (vertical dashed pink lines).
    We show the difference in logarithmic space between the pseudo-evolution of the mass and concentration.
    In the first panel, the shaded regions show the 68th percentile of the absolute deviation from pseudo-evolution as a function of time, for the entire random sample, in the time ranges of interest for the mass and concentration respectively.}
    \label{fig:scatter}
\end{figure*}

It is tempting to synthesise the present-day concentration from the full mass assembly history, by superposing the effect of each merger event upon pseudo-evolution.
However, we show in \autoref{fig:scatter} that even small deviations from pseudo-evolution in mass can cause large fluctuations in concentration.
This sensitivity of the concentration to small mergers and the stochastic nature of mergers make it virtually impossible to predict the concentration of an individual halo from its formation history without some uncertainty.
In \autoref{fig:scatter}, we select the five haloes from the random catalogue that have the most quiescent mass assembly histories in the last five dynamical times.
We do this by minimising the deviation of the mass assembly history from pure pseudo-evolution.
We calculate the forward pseudo-evolution of mass from the halo properties at $T(a;a_{\rm ref}=1)=-5$, which is marked by the vertical dotted dark blue lines in \autoref{fig:scatter}, and quantify the deviation in mass evolution by $\sum |M_{\rm hist}/M_{\rm pseudo}-1|$, where $M_{\rm hist}$ is the actual evolution, $M_{\rm pseudo}$ is the forward pseudo-evolution for each halo, and the sum is taken over the available snapshots in the last five dynamical times.
The dark blue lines in the figure show the logarithmic deviation in mass.
For these haloes with quiescent mass assembly histories, we then compare between the actual and pseudo-evolution of concentration during the last two dynamical times, since $T(a;a_{\rm ref}=1)=-2$ (vertical dashed pink lines), to exclude the effect of mass evolution before the controlled period.
The comparison of concentrations is shown as pink lines in \autoref{fig:scatter}.
In the first panel, we also show the 68\% range of the absolute deviation from both mass and concentration pseudo-evolutions for the entire random sample.

From the figure it is apparent that even selecting the 
most quiescent haloes in our sample, 
which are usually considered relaxed, does not greatly 
reduce fluctuations in concentration.
This shows that even very minor mass accretion can 
affect halo structures. The fluctuations in concentration seen in 
\autoref{fig:scatter} could also be partly due to the 
finite number of snapshots available from the simulation, 
which leaves events that happen between the 
discrete snapshots undetected.
We also notice that for some haloes (e.g., Halos 2 and 3), the concentration evolution has a general trend that deviates from the pseudo-evolution prediction. This is likely due to the oversimplified assumptions in the pseudo-evolution model, such as an NFW profile and an isolated halo, which may not hold true in simulations \citep[e.g.,][]{diemer_kravtsov14}.  
The environments of individual haloes and further details of mergers, such as the relative 
velocities of the progenitors, the exact orbit of the 
incoming object, 
the detailed density profiles of each progenitor, 
are beyond the scope of this work, 
and might also have caused part of 
the uncertainty that we observe.

At this point we reflect on the limited ability to predict halo concentration with a linear regression of the mass assembly history in \autoref{sec:lin_comb}, and conclude that this is unsurprising, because a fixed set of linear coefficients is naturally incapable of describing a convolution of merger responses at different times.
Also, in the top panel of \autoref{fig:spearmanr}, the concentration of the cluster-size halo sample is less correlated with the step-wise mass assembly history than for the less massive samples, probably ascribable to its higher frequency of merger events.
On the other hand, the higher frequency of mergers in the cluster-size sample causes its stronger anticorrelation between the concentration and the mass increment at late times in the bottom panel of \autoref{fig:spearmanr}.
The oscillatory behaviour in the bottom panel of \autoref{fig:spearmanr} has approximately the same timescales as the oscillation of the concentration and scale radius in \autoref{fig:unitime}, and it is now apparent that it arises from merger responses.

We have demonstrated that mergers play a vital role in shaping the internal structures of haloes, and merger events that happen at different epochs trigger responses with nontrivial forms, preventing a simple description of the combined end result, and contribute to the scatter in the scale radius and hence concentration.

\section{Discussion and Summary}
\label{sec:summary}

In this study, we investigate the connection between halo concentration and halo mass assembly history using the halo catalogues and merger trees from the Dark Sky Simulations. 
In particular, we scrutinise the effect of mergers on the subsequent evolution of halo concentration.
We summarise our primary results as follows:

\begin{itemize}
\item Conventionally defined halo formation times, such as the scale factor at which a halo reaches 50\% of its contemporary mass, exhibit significant correlations with the present-day halo concentration. In fact, the same holds true for the broad range of mass fractions between approximately 30\% and 70\%. 
A linear combination of $a(m_i)$, where $m_i$'s are different choices of mass fractions, correlates with present-day concentration better than any individual $a(m_i)$, but still does not fully account for the scatter in concentration at a fixed halo mass. The same conclusions apply when we use the mass fractions at different times, $m(a)$, instead of $a(m)$. For more details, see \autoref{fig:spearmanr}.\\
\item Major mergers induce dramatic 
changes to halo concentrations. 
These responses linger over a period of several 
dynamical times, corresponding to many Gyr.
The evolution of concentration due to a merger 
can be associated with the orbital dynamics of the 
merger and is largely universal.
Minor mergers have similar, but less dramatic 
effects on concentration compared with major mergers.
In the absence of merger events, 
pseudo-evolution causes a gradual increase in halo concentration and halo mass (\autoref{fig:M-c-R}), 
in agreement with \citet{diemer_etal13}. See \autoref{fig:orbit}, \autoref{fig:unitime}, and \autoref{fig:minor}.\\
\item The cumulative effect of major mergers
and frequent minor mergers leads to a scatter 
in concentration at fixed halo mass and 
fixed formation time (any conventional definition).
At fixed halo mass, the scatter can be reduced from 0.2 dex to below 0.1 dex when we control for both formation time and merger events.
Even minor mergers impart non-negligible 
alterations to concentrations.
Haloes with quiescent mass assembly histories 
experience fewer fluctuations in concentration, 
but still with an irreducible scatter, 
due to unresolved small mergers.
See \autoref{fig:sigmalogc} and \autoref{fig:scatter}.
\end{itemize}

In this work, we have developed a further understanding of the relation between halo concentrations and mass assembly histories.
Our results show that the correlation strengths with concentration at multiple intermediate epochs of the assembly history are similar and relatively high, 
in accord with previously found 
concentration--formation time relations \citep[e.g.,][]{wechsler02,zhao_etal09}. 
Our findings support the use of the half-mass scale, 
$a_{1/2}$, as an effective definition of formation time, whereas a variety of similar formation time definitions 
would yield similar insight into concentrations. 
However, we also argue that such simple characterisations 
of the mass assembly history inevitably omit information 
on halo structure and leave a non-negligible residual scatter.

We find that merger events during the assembly of haloes 
contribute to the scatter in the 
concentration--formation time relation (at fixed halo mass), 
as was suggested by the results of, e.g., 
\citet{wechsler02, maccio08}, and \citet{Rey_etal2019}.
We broaden the discussion of the impact of recent mergers 
on the measurement of concentration in \citet{ludlow2012}, 
confirming their explanation of the features in the merger 
response with a case study of the orbital processes during 
a merger,
and these fluctuations in concentration induced by mergers also lead to the non-monotonic relation between concentration and formation time observed by \citet{ludlow2012}.
We recognise the significant effect of mergers 
on halo concentrations, which greatly exceeds the secular 
evolution during quiescent periods. The effect of mergers 
lasts for several Gyr (a few dynamical times). Our results 
also establish the universality of halo responses 
to merger events.

These results can have important implications for the interpretation of observations, as the observed density profiles of the dark component of clusters are systematically dependent on the merger history.
The concentration--mass relation is broadly adopted for inferring concentrations from mass measurements, comparing measured concentrations against theoretical predictions, or modelling other halo properties with concentrations \citep[e.g.,][]{comerford2007,duffy2008,mandelbaum2008,bhattacharya2013,CLASH2015,Li2019SPARC}.
Based on our findings regarding the scatter around the mean relation due to mergers, we advise caution in the application of the concentration--mass or concentration--formation time relation without taking these effects into account.

Our study provides insight into secondary halo biases, 
commonly known as halo assembly bias 
\citep[e.g.,][]{gao_etal05,li_etal08,mao_etal18}, 
the dependence of halo clustering on halo properties 
other than mass. \citet{wechsler06} first found that 
with fixed masses above the typical collapse mass, 
haloes with lower concentrations cluster more strongly 
than haloes with higher concentrations.
\citet{lee2017}, for example, found that the scale radii of haloes evolve very differently in regions of different environmental densities.
Our findings suggest that these are primarily due 
to the suppression of concentration by merger events, 
which happen more frequently in denser environments.
We expect similar coupling of the environmental preference 
of mergers and the impact of mergers on 
other secondary halo properties to be present.

Our analyses are performed at the halo level, 
which introduces a dependence on the halo finding algorithm.
We limit our characterisation of the mass assembly history 
to linear descriptions, and do not propose a mathematical 
model of the concentration.
We are also unable to resolve all merger events, 
and further details, including the initial profiles, 
initial velocities, and trajectories of merging objects, 
are beyond the scope of this work.
Each of these important issues merits further study.
More sophisticated statistics or machine learning techniques 
might be more effective in extracting information on 
concentrations from assembly histories.
Using explicit mathematical descriptions of concentration responses to mergers, together with a comprehensive demographic study of merger events with even higher mass and temporal resolutions is a possible way of improving predictions of concentrations.
We are hopeful that such follow-up studies could greatly enhance our understanding of halo formation and structure.

\section*{Acknowledgements}
The authors thank Susmita Adhikari, Rachel Bezanson, Andrew Hearin, Arthur Kosowsky, Martin Rey, and Qiong Zhang for useful discussions.
This research made use part of the Dark Sky Simulations (\http{darksky.slac.stanford.edu}) produced using an INCITE 2014 allocation on the Oak Ridge Leadership Computing Facility at Oak Ridge National Laboratory, a U.S.\ Department of Energy Office; the authors thank Sam Skillman, Mike Warren, Matt Turk, and other Dark Sky collaborators for their efforts in creating these simulations and for providing access to them. 
This research made use of computational resources at SLAC National Accelerator Laboratory, a U.S.\ Department of Energy Office; YYM and RHW thank the support of the SLAC computational team. 
KW and ARZ are supported by the US National Science Foundation (NSF) through grant AST 1517563 and by the Pittsburgh Particle Physics Astrophysics and Cosmology Center (PITT PACC). 
Support for YYM\ was provided by the Pittsburgh Particle Physics, Astrophysics and Cosmology Center through the Samuel P.\ Langley PITT PACC Postdoctoral Fellowship, and by NASA through the NASA Hubble Fellowship grant no.\ HST-HF2-51441.001 awarded by the Space Telescope Science Institute, which is operated by the Association of Universities for Research in Astronomy, Incorporated, under NASA contract NAS5-26555. 
FvdB is supported by the US National Science Foundation through grant AST 1516962. FvdB received additional support from the Klaus Tschira foundation, and from the National Aeronautics and Space Administration through Grants 17-ATP17-0028 and 19-ATP19-0059 issued as part of the Astrophysics Theory Program. 

This research made use of Python, along with many community-developed or maintained software packages, including
IPython \citep{ipython},
Jupyter (\https{jupyter.org}),
Matplotlib \citep{matplotlib},
NumPy \citep{numpy},
SciPy \citep{scipy},
Astropy \citep{astropy},
Pandas \citep{pandas},
and Scikit-learn \citep{scikit-learn}.
This research made use of NASA's Astrophysics Data System for bibliographic information.



\bibliographystyle{mnras}
\bibliography{main} 

\begin{thebibliography}{}
\makeatletter
\relax
\def\mn@urlcharsother{\let\do\@makeother \do\$\do\&\do\#\do\^\do\_\do\%\do\~}
\def\mn@doi{\begingroup\mn@urlcharsother \@ifnextchar [ {\mn@doi@}
  {\mn@doi@[]}}
\def\mn@doi@[#1]#2{\def\@tempa{#1}\ifx\@tempa\@empty \href
  {http://dx.doi.org/#2} {doi:#2}\else \href {http://dx.doi.org/#2} {#1}\fi
  \endgroup}
\def\mn@eprint#1#2{\mn@eprint@#1:#2::\@nil}
\def\mn@eprint@arXiv#1{\href {http://arxiv.org/abs/#1} {{\tt arXiv:#1}}}
\def\mn@eprint@dblp#1{\href {http://dblp.uni-trier.de/rec/bibtex/#1.xml}
  {dblp:#1}}
\def\mn@eprint@#1:#2:#3:#4\@nil{\def\@tempa {#1}\def\@tempb {#2}\def\@tempc
  {#3}\ifx \@tempc \@empty \let \@tempc \@tempb \let \@tempb \@tempa \fi \ifx
  \@tempb \@empty \def\@tempb {arXiv}\fi \@ifundefined
  {mn@eprint@\@tempb}{\@tempb:\@tempc}{\expandafter \expandafter \csname
  mn@eprint@\@tempb\endcsname \expandafter{\@tempc}}}

\bibitem[\protect\citeauthoryear{{Abbott} et~al.,}{{Abbott}
  et~al.}{2019}]{DES2018}
{Abbott} T.~M.~C.,  et~al., 2019, \mn@doi [\prd] {10.1103/PhysRevD.99.123505},
  \href {https://ui.adsabs.harvard.edu/abs/2019PhRvD..99l3505A} {99, 123505}

\bibitem[\protect\citeauthoryear{{Astropy Collaboration} et~al.,}{{Astropy
  Collaboration} et~al.}{2013}]{astropy}
{Astropy Collaboration} et~al., 2013, \mn@doi [\aap]
  {10.1051/0004-6361/201322068}, \href
  {http://adsabs.harvard.edu/abs/2013A%26A...558A..33A} {558, A33}

\bibitem[\protect\citeauthoryear{{Behroozi}, {Wechsler}  \& {Wu}}{{Behroozi}
  et~al.}{2013a}]{rockstar}
{Behroozi} P.~S.,  {Wechsler} R.~H.,   {Wu} H.-Y.,  2013a, \mn@doi [\apj]
  {10.1088/0004-637X/762/2/109}, \href
  {https://ui.adsabs.harvard.edu/abs/2013ApJ...762..109B} {762, 109}

\bibitem[\protect\citeauthoryear{{Behroozi}, {Wechsler}, {Wu}, {Busha},
  {Klypin}  \& {Primack}}{{Behroozi} et~al.}{2013b}]{behroozi_trees13}
{Behroozi} P.~S.,  {Wechsler} R.~H.,  {Wu} H.-Y.,  {Busha} M.~T.,  {Klypin}
  A.~A.,   {Primack} J.~R.,  2013b, \mn@doi [\apj]
  {10.1088/0004-637X/763/1/18}, \href
  {https://ui.adsabs.harvard.edu/abs/2013ApJ...763...18B} {763, 18}

\bibitem[\protect\citeauthoryear{{Benson}, {Ludlow}  \& {Cole}}{{Benson}
  et~al.}{2019}]{benson2019}
{Benson} A.~J.,  {Ludlow} A.,   {Cole} S.,  2019, \mn@doi [\mnras]
  {10.1093/mnras/stz695}, \href
  {https://ui.adsabs.harvard.edu/abs/2019MNRAS.485.5010B} {485, 5010}

\bibitem[\protect\citeauthoryear{{Berlind} \& {Weinberg}}{{Berlind} \&
  {Weinberg}}{2002}]{berlind02}
{Berlind} A.~A.,  {Weinberg} D.~H.,  2002, \mn@doi [\apj] {10.1086/341469},
  \href {http://adsabs.harvard.edu/abs/2002ApJ...575..587B} {575, 587}

\bibitem[\protect\citeauthoryear{{Bhattacharya}, {Habib}, {Heitmann}  \&
  {Vikhlinin}}{{Bhattacharya} et~al.}{2013}]{bhattacharya2013}
{Bhattacharya} S.,  {Habib} S.,  {Heitmann} K.,   {Vikhlinin} A.,  2013,
  \mn@doi [\apj] {10.1088/0004-637X/766/1/32}, \href
  {https://ui.adsabs.harvard.edu/abs/2013ApJ...766...32B} {766, 32}

\bibitem[\protect\citeauthoryear{{Blumenthal}, {Faber}, {Primack}  \&
  {Rees}}{{Blumenthal} et~al.}{1984}]{blumenthal_etal84}
{Blumenthal} G.~R.,  {Faber} S.~M.,  {Primack} J.~R.,   {Rees} M.~J.,  1984,
  \mn@doi [\nat] {10.1038/311517a0}, \href
  {https://ui.adsabs.harvard.edu/abs/1984Natur.311..517B} {311, 517}

\bibitem[\protect\citeauthoryear{{Bond}, {Cole}, {Efstathiou}  \&
  {Kaiser}}{{Bond} et~al.}{1991}]{Bond91}
{Bond} J.~R.,  {Cole} S.,  {Efstathiou} G.,   {Kaiser} N.,  1991, \mn@doi
  [\apj] {10.1086/170520}, \href
  {https://ui.adsabs.harvard.edu/abs/1991ApJ...379..440B} {379, 440}

\bibitem[\protect\citeauthoryear{{Bryan} \& {Norman}}{{Bryan} \&
  {Norman}}{1998}]{bryan_norman98}
{Bryan} G.~L.,  {Norman} M.~L.,  1998, \mn@doi [\apj] {10.1086/305262}, \href
  {https://ui.adsabs.harvard.edu/abs/1998ApJ...495...80B} {495, 80}

\bibitem[\protect\citeauthoryear{{Bullock}, {Kolatt}, {Sigad}, {Somerville},
  {Kravtsov}, {Klypin}, {Primack}  \& {Dekel}}{{Bullock}
  et~al.}{2001}]{bullock01}
{Bullock} J.~S.,  {Kolatt} T.~S.,  {Sigad} Y.,  {Somerville} R.~S.,  {Kravtsov}
  A.~V.,  {Klypin} A.~A.,  {Primack} J.~R.,   {Dekel} A.,  2001, \mn@doi
  [\mnras] {10.1046/j.1365-8711.2001.04068.x}, \href
  {https://ui.adsabs.harvard.edu/abs/2001MNRAS.321..559B} {321, 559}

\bibitem[\protect\citeauthoryear{{Child}, {Habib}, {Heitmann}, {Frontiere},
  {Finkel}, {Pope}  \& {Morozov}}{{Child} et~al.}{2018}]{child2018}
{Child} H.~L.,  {Habib} S.,  {Heitmann} K.,  {Frontiere} N.,  {Finkel} H.,
  {Pope} A.,   {Morozov} V.,  2018, \mn@doi [\apj] {10.3847/1538-4357/aabf95},
  \href {https://ui.adsabs.harvard.edu/abs/2018ApJ...859...55C} {859, 55}

\bibitem[\protect\citeauthoryear{{Comerford} \& {Natarajan}}{{Comerford} \&
  {Natarajan}}{2007}]{comerford2007}
{Comerford} J.~M.,  {Natarajan} P.,  2007, \mn@doi [\mnras]
  {10.1111/j.1365-2966.2007.11934.x}, \href
  {https://ui.adsabs.harvard.edu/abs/2007MNRAS.379..190C} {379, 190}

\bibitem[\protect\citeauthoryear{{Cooray} \& {Sheth}}{{Cooray} \&
  {Sheth}}{2002}]{cooray02}
{Cooray} A.,  {Sheth} R.,  2002, \mn@doi [\physrep]
  {10.1016/S0370-1573(02)00276-4}, \href
  {http://adsabs.harvard.edu/abs/2002PhR...372....1C} {372, 1}

\bibitem[\protect\citeauthoryear{{Correa}, {Wyithe}, {Schaye}  \&
  {Duffy}}{{Correa} et~al.}{2015a}]{Correa2015I}
{Correa} C.~A.,  {Wyithe} J. S.~B.,  {Schaye} J.,   {Duffy} A.~R.,  2015a,
  \mn@doi [\mnras] {10.1093/mnras/stv689}, \href
  {https://ui.adsabs.harvard.edu/abs/2015MNRAS.450.1514C} {450, 1514}

\bibitem[\protect\citeauthoryear{{Correa}, {Wyithe}, {Schaye}  \&
  {Duffy}}{{Correa} et~al.}{2015b}]{Correa2015II}
{Correa} C.~A.,  {Wyithe} J. S.~B.,  {Schaye} J.,   {Duffy} A.~R.,  2015b,
  \mn@doi [\mnras] {10.1093/mnras/stv697}, \href
  {https://ui.adsabs.harvard.edu/abs/2015MNRAS.450.1521C} {450, 1521}

\bibitem[\protect\citeauthoryear{{Correa}, {Wyithe}, {Schaye}  \&
  {Duffy}}{{Correa} et~al.}{2015c}]{Correa2015III}
{Correa} C.~A.,  {Wyithe} J. S.~B.,  {Schaye} J.,   {Duffy} A.~R.,  2015c,
  \mn@doi [\mnras] {10.1093/mnras/stv1363}, \href
  {https://ui.adsabs.harvard.edu/abs/2015MNRAS.452.1217C} {452, 1217}

\bibitem[\protect\citeauthoryear{{DeRose} et~al.,}{{DeRose}
  et~al.}{2019}]{deRose2019_aemulus}
{DeRose} J.,  et~al., 2019, \mn@doi [\apj] {10.3847/1538-4357/ab1085}, \href
  {https://ui.adsabs.harvard.edu/abs/2019ApJ...875...69D} {875, 69}

\bibitem[\protect\citeauthoryear{{Diemer}}{{Diemer}}{2018}]{colossus2018}
{Diemer} B.,  2018, \mn@doi [\apjs] {10.3847/1538-4365/aaee8c}, \href
  {https://ui.adsabs.harvard.edu/abs/2018ApJS..239...35D} {239, 35}

\bibitem[\protect\citeauthoryear{{Diemer} \& {Kravtsov}}{{Diemer} \&
  {Kravtsov}}{2014}]{diemer_kravtsov14}
{Diemer} B.,  {Kravtsov} A.~V.,  2014, \mn@doi [\apj]
  {10.1088/0004-637X/789/1/1}, \href
  {http://adsabs.harvard.edu/abs/2014ApJ...789....1D} {789, 1}

\bibitem[\protect\citeauthoryear{{Diemer} \& {Kravtsov}}{{Diemer} \&
  {Kravtsov}}{2015}]{diemer2015}
{Diemer} B.,  {Kravtsov} A.~V.,  2015, \mn@doi [\apj]
  {10.1088/0004-637X/799/1/108}, \href
  {https://ui.adsabs.harvard.edu/abs/2015ApJ...799..108D} {799, 108}

\bibitem[\protect\citeauthoryear{{Diemer}, {More}  \& {Kravtsov}}{{Diemer}
  et~al.}{2013}]{diemer_etal13}
{Diemer} B.,  {More} S.,   {Kravtsov} A.~V.,  2013, \mn@doi [ApJ]
  {10.1088/0004-637X/766/1/25}, \href
  {http://adsabs.harvard.edu/abs/2013ApJ...766...25D} {766, 25}

\bibitem[\protect\citeauthoryear{{Duffy}, {Schaye}, {Kay}  \& {Dalla
  Vecchia}}{{Duffy} et~al.}{2008}]{duffy2008}
{Duffy} A.~R.,  {Schaye} J.,  {Kay} S.~T.,   {Dalla Vecchia} C.,  2008, \mn@doi
  [\mnras] {10.1111/j.1745-3933.2008.00537.x}, \href
  {https://ui.adsabs.harvard.edu/abs/2008MNRAS.390L..64D} {390, L64}

\bibitem[\protect\citeauthoryear{{Einasto}}{{Einasto}}{1965}]{einasto65}
{Einasto} J.,  1965, Trudy Inst. Astrofiz. Alma-Ata 51, 87

\bibitem[\protect\citeauthoryear{{Gao}, {Springel}  \& {White}}{{Gao}
  et~al.}{2005}]{gao_etal05}
{Gao} L.,  {Springel} V.,   {White} S. D.~M.,  2005, \mn@doi [\mnras]
  {10.1111/j.1745-3933.2005.00084.x}, \href
  {https://ui.adsabs.harvard.edu/abs/2005MNRAS.363L..66G} {363, L66}

\bibitem[\protect\citeauthoryear{{Gao}, {Navarro}, {Cole}, {Frenk}, {White},
  {Springel}, {Jenkins}  \& {Neto}}{{Gao} et~al.}{2008}]{gao08}
{Gao} L.,  {Navarro} J.~F.,  {Cole} S.,  {Frenk} C.~S.,  {White} S.~D.~M.,
  {Springel} V.,  {Jenkins} A.,   {Neto} A.~F.,  2008, \mn@doi [\mnras]
  {10.1111/j.1365-2966.2008.13277.x}, \href
  {http://adsabs.harvard.edu/abs/2008MNRAS.387..536G} {387, 536}

\bibitem[\protect\citeauthoryear{{Heitmann} et~al.,}{{Heitmann}
  et~al.}{2019}]{heitmann_2019_outerrim}
{Heitmann} K.,  et~al., 2019, \mn@doi [\apjs] {10.3847/1538-4365/ab4da1}, \href
  {https://ui.adsabs.harvard.edu/abs/2019ApJS..245...16H} {245, 16}

\bibitem[\protect\citeauthoryear{Hunter}{Hunter}{2007}]{matplotlib}
Hunter J.~D.,  2007, \mn@doi [Computing in Science Engineering]
  {10.1109/MCSE.2007.55}, 9, 90

\bibitem[\protect\citeauthoryear{{Jiang} \& {van den Bosch}}{{Jiang} \& {van
  den Bosch}}{2016}]{Jiang_vdB16}
{Jiang} F.,  {van den Bosch} F.~C.,  2016, \mn@doi [\mnras]
  {10.1093/mnras/stw439}, \href
  {https://ui.adsabs.harvard.edu/abs/2016MNRAS.458.2848J} {458, 2848}

\bibitem[\protect\citeauthoryear{{Johnson}, {Benson}  \& {Grin}}{{Johnson}
  et~al.}{2020}]{Johnson2020}
{Johnson} T.,  {Benson} A.~J.,   {Grin} D.,  2020, arXiv e-prints, \href
  {https://ui.adsabs.harvard.edu/abs/2020arXiv200615231J} {p. arXiv:2006.15231}

\bibitem[\protect\citeauthoryear{Jones, Oliphant, Peterson  et~al.}{Jones
  et~al.}{2001}]{scipy}
Jones E.,  Oliphant T.,  Peterson P.,   et~al., 2001, {SciPy}: Open source
  scientific tools for {Python}, \url {http://www.scipy.org/}

\bibitem[\protect\citeauthoryear{{Klypin}, {Trujillo-Gomez}  \&
  {Primack}}{{Klypin} et~al.}{2011a}]{bolshoi_11}
{Klypin} A.~A.,  {Trujillo-Gomez} S.,   {Primack} J.,  2011a, \mn@doi [\apj]
  {10.1088/0004-637X/740/2/102}, \href
  {https://ui.adsabs.harvard.edu/abs/2011ApJ...740..102K} {740, 102}

\bibitem[\protect\citeauthoryear{{Klypin}, {Trujillo-Gomez}  \&
  {Primack}}{{Klypin} et~al.}{2011b}]{klypin11_rs}
{Klypin} A.~A.,  {Trujillo-Gomez} S.,   {Primack} J.,  2011b, \mn@doi [\apj]
  {10.1088/0004-637X/740/2/102}, \href
  {https://ui.adsabs.harvard.edu/abs/2011ApJ...740..102K} {740, 102}

\bibitem[\protect\citeauthoryear{{Klypin}, {Yepes}, {Gottl{\"o}ber}, {Prada}
  \& {He{\ss}}}{{Klypin} et~al.}{2016}]{klypin_etal16}
{Klypin} A.,  {Yepes} G.,  {Gottl{\"o}ber} S.,  {Prada} F.,   {He{\ss}} S.,
  2016, \mn@doi [\mnras] {10.1093/mnras/stw248}, \href
  {https://ui.adsabs.harvard.edu/abs/2016MNRAS.457.4340K} {457, 4340}

\bibitem[\protect\citeauthoryear{{Komatsu} et~al.,}{{Komatsu}
  et~al.}{2011}]{komatsu_WMAP711}
{Komatsu} E.,  et~al., 2011, \mn@doi [\apjs] {10.1088/0067-0049/192/2/18},
  \href {https://ui.adsabs.harvard.edu/abs/2011ApJS..192...18K} {192, 18}

\bibitem[\protect\citeauthoryear{{Lee}, {Primack}, {Behroozi},
  {Rodr{\'\i}guez-Puebla}, {Hellinger}  \& {Dekel}}{{Lee}
  et~al.}{2017}]{lee2017}
{Lee} C.~T.,  {Primack} J.~R.,  {Behroozi} P.,  {Rodr{\'\i}guez-Puebla} A.,
  {Hellinger} D.,   {Dekel} A.,  2017, \mn@doi [\mnras]
  {10.1093/mnras/stw3348}, \href
  {https://ui.adsabs.harvard.edu/abs/2017MNRAS.466.3834L} {466, 3834}

\bibitem[\protect\citeauthoryear{{Lee}, {Primack}, {Behroozi},
  {Rodr{\'\i}guez-Puebla}, {Hellinger}  \& {Dekel}}{{Lee}
  et~al.}{2018}]{lee2018}
{Lee} C.~T.,  {Primack} J.~R.,  {Behroozi} P.,  {Rodr{\'\i}guez-Puebla} A.,
  {Hellinger} D.,   {Dekel} A.,  2018, \mn@doi [\mnras]
  {10.1093/mnras/sty2538}, \href
  {https://ui.adsabs.harvard.edu/abs/2018MNRAS.481.4038L} {481, 4038}

\bibitem[\protect\citeauthoryear{{Li}, {Mo}, {van den Bosch}  \& {Lin}}{{Li}
  et~al.}{2007}]{li_mo_vdb_07}
{Li} Y.,  {Mo} H.~J.,  {van den Bosch} F.~C.,   {Lin} W.~P.,  2007, \mn@doi
  [\mnras] {10.1111/j.1365-2966.2007.11942.x}, \href
  {https://ui.adsabs.harvard.edu/abs/2007MNRAS.379..689L} {379, 689}

\bibitem[\protect\citeauthoryear{{Li}, {Mo}  \& {Gao}}{{Li}
  et~al.}{2008}]{li_etal08}
{Li} Y.,  {Mo} H.~J.,   {Gao} L.,  2008, \mn@doi [\mnras]
  {10.1111/j.1365-2966.2008.13667.x}, \href
  {https://ui.adsabs.harvard.edu/abs/2008MNRAS.389.1419L} {389, 1419}

\bibitem[\protect\citeauthoryear{{Li}, {Lelli}, {McGaugh}, {Starkman}  \&
  {Schombert}}{{Li} et~al.}{2019}]{Li2019SPARC}
{Li} P.,  {Lelli} F.,  {McGaugh} S.~S.,  {Starkman} N.,   {Schombert} J.~M.,
  2019, \mn@doi [\mnras] {10.1093/mnras/sty2968}, \href
  {https://ui.adsabs.harvard.edu/abs/2019MNRAS.482.5106L} {482, 5106}

\bibitem[\protect\citeauthoryear{{Ludlow}, {Navarro}, {Li}, {Angulo},
  {Boylan-Kolchin}  \& {Bett}}{{Ludlow} et~al.}{2012}]{ludlow2012}
{Ludlow} A.~D.,  {Navarro} J.~F.,  {Li} M.,  {Angulo} R.~E.,  {Boylan-Kolchin}
  M.,   {Bett} P.~E.,  2012, \mn@doi [\mnras]
  {10.1111/j.1365-2966.2012.21892.x}, \href
  {https://ui.adsabs.harvard.edu/abs/2012MNRAS.427.1322L} {427, 1322}

\bibitem[\protect\citeauthoryear{{Ludlow} et~al.,}{{Ludlow}
  et~al.}{2013}]{ludlow_etal13}
{Ludlow} A.~D.,  et~al., 2013, \mn@doi [\mnras] {10.1093/mnras/stt526}, \href
  {http://adsabs.harvard.edu/abs/2013MNRAS.432.1103L} {432, 1103}

\bibitem[\protect\citeauthoryear{{Ludlow}, {Navarro}, {Angulo},
  {Boylan-Kolchin}, {Springel}, {Frenk}  \& {White}}{{Ludlow}
  et~al.}{2014}]{ludlow2014}
{Ludlow} A.~D.,  {Navarro} J.~F.,  {Angulo} R.~E.,  {Boylan-Kolchin} M.,
  {Springel} V.,  {Frenk} C.,   {White} S. D.~M.,  2014, \mn@doi [\mnras]
  {10.1093/mnras/stu483}, \href
  {https://ui.adsabs.harvard.edu/abs/2014MNRAS.441..378L} {441, 378}

\bibitem[\protect\citeauthoryear{{Ludlow}, {Bose}, {Angulo}, {Wang},
  {Hellwing}, {Navarro}, {Cole}  \& {Frenk}}{{Ludlow}
  et~al.}{2016}]{ludlow2016}
{Ludlow} A.~D.,  {Bose} S.,  {Angulo} R.~E.,  {Wang} L.,  {Hellwing} W.~A.,
  {Navarro} J.~F.,  {Cole} S.,   {Frenk} C.~S.,  2016, \mn@doi [\mnras]
  {10.1093/mnras/stw1046}, \href
  {https://ui.adsabs.harvard.edu/abs/2016MNRAS.460.1214L} {460, 1214}

\bibitem[\protect\citeauthoryear{{Macci{\`o}}, {Dutton}  \& {van den
  Bosch}}{{Macci{\`o}} et~al.}{2008}]{maccio08}
{Macci{\`o}} A.~V.,  {Dutton} A.~A.,   {van den Bosch} F.~C.,  2008, \mn@doi
  [\mnras] {10.1111/j.1365-2966.2008.14029.x}, \href
  {https://ui.adsabs.harvard.edu/abs/2008MNRAS.391.1940M} {391, 1940}

\bibitem[\protect\citeauthoryear{{Mandelbaum}, {Seljak}  \&
  {Hirata}}{{Mandelbaum} et~al.}{2008}]{mandelbaum2008}
{Mandelbaum} R.,  {Seljak} U.,   {Hirata} C.~M.,  2008, \mn@doi [\jcap]
  {10.1088/1475-7516/2008/08/006}, \href
  {https://ui.adsabs.harvard.edu/abs/2008JCAP...08..006M} {2008, 006}

\bibitem[\protect\citeauthoryear{{Mao}, {Zentner}  \& {Wechsler}}{{Mao}
  et~al.}{2018}]{mao_etal18}
{Mao} Y.-Y.,  {Zentner} A.~R.,   {Wechsler} R.~H.,  2018, \mn@doi [\mnras]
  {10.1093/mnras/stx3111}, \href
  {https://ui.adsabs.harvard.edu/abs/2018MNRAS.474.5143M} {474, 5143}

\bibitem[\protect\citeauthoryear{{McBride}, {Fakhouri}  \& {Ma}}{{McBride}
  et~al.}{2009}]{McBride2009}
{McBride} J.,  {Fakhouri} O.,   {Ma} C.-P.,  2009, \mn@doi [\mnras]
  {10.1111/j.1365-2966.2009.15329.x}, \href
  {https://ui.adsabs.harvard.edu/abs/2009MNRAS.398.1858M} {398, 1858}

\bibitem[\protect\citeauthoryear{McKinney}{McKinney}{2010}]{pandas}
McKinney W.,  2010, in van~der Walt S.,  Millman J.,  eds, Proceedings of the
  9th Python in Science Conference. pp 51 -- 56

\bibitem[\protect\citeauthoryear{{Merten} et~al.,}{{Merten}
  et~al.}{2015}]{CLASH2015}
{Merten} J.,  et~al., 2015, \mn@doi [\apj] {10.1088/0004-637X/806/1/4}, \href
  {https://ui.adsabs.harvard.edu/abs/2015ApJ...806....4M} {806, 4}

\bibitem[\protect\citeauthoryear{{Moore}, {Governato}, {Quinn}, {Stadel}  \&
  {Lake}}{{Moore} et~al.}{1998}]{moore98_profile}
{Moore} B.,  {Governato} F.,  {Quinn} T.,  {Stadel} J.,   {Lake} G.,  1998,
  \mn@doi [\apjl] {10.1086/311333}, \href
  {https://ui.adsabs.harvard.edu/abs/1998ApJ...499L...5M} {499, L5}

\bibitem[\protect\citeauthoryear{{Navarro}, {Frenk}  \& {White}}{{Navarro}
  et~al.}{1995}]{nfw95}
{Navarro} J.~F.,  {Frenk} C.~S.,   {White} S. D.~M.,  1995, \mn@doi [\mnras]
  {10.1093/mnras/275.1.56}, \href
  {https://ui.adsabs.harvard.edu/abs/1995MNRAS.275...56N} {275, 56}

\bibitem[\protect\citeauthoryear{{Navarro}, {Frenk}  \& {White}}{{Navarro}
  et~al.}{1996}]{nfw96}
{Navarro} J.~F.,  {Frenk} C.~S.,   {White} S. D.~M.,  1996, \mn@doi [\apj]
  {10.1086/177173}, \href
  {https://ui.adsabs.harvard.edu/abs/1996ApJ...462..563N} {462, 563}

\bibitem[\protect\citeauthoryear{{Navarro}, {Frenk}  \& {White}}{{Navarro}
  et~al.}{1997}]{nfw97}
{Navarro} J.~F.,  {Frenk} C.~S.,   {White} S. D.~M.,  1997, \mn@doi [\apj]
  {10.1086/304888}, \href
  {https://ui.adsabs.harvard.edu/abs/1997ApJ...490..493N} {490, 493}

\bibitem[\protect\citeauthoryear{{Navarro} et~al.,}{{Navarro}
  et~al.}{2010}]{navarro10}
{Navarro} J.~F.,  et~al., 2010, \mn@doi [\mnras]
  {10.1111/j.1365-2966.2009.15878.x}, \href
  {http://adsabs.harvard.edu/abs/2010MNRAS.402...21N} {402, 21}

\bibitem[\protect\citeauthoryear{{Neto} et~al.,}{{Neto} et~al.}{2007}]{neto07}
{Neto} A.~F.,  et~al., 2007, \mn@doi [\mnras]
  {10.1111/j.1365-2966.2007.12381.x}, \href
  {https://ui.adsabs.harvard.edu/abs/2007MNRAS.381.1450N} {381, 1450}

\bibitem[\protect\citeauthoryear{Pedregosa et~al.,}{Pedregosa
  et~al.}{2011}]{scikit-learn}
Pedregosa F.,  et~al., 2011, Journal of Machine Learning Research

\bibitem[\protect\citeauthoryear{P\'erez \& Granger}{P\'erez \&
  Granger}{2007}]{ipython}
P\'erez F.,  Granger B.~E.,  2007, \mn@doi [Computing in Science Engineering]
  {10.1109/MCSE.2007.53}, 9, 21

\bibitem[\protect\citeauthoryear{{Planck Collaboration} et~al.,}{{Planck
  Collaboration} et~al.}{2014}]{planck13}
{Planck Collaboration} et~al., 2014, \mn@doi [\aap]
  {10.1051/0004-6361/201321591}, \href
  {https://ui.adsabs.harvard.edu/abs/2014A&A...571A..16P} {571, A16}

\bibitem[\protect\citeauthoryear{{Planck Collaboration} et~al.,}{{Planck
  Collaboration} et~al.}{2018}]{Planck2018}
{Planck Collaboration} et~al., 2018, arXiv e-prints, \href
  {https://ui.adsabs.harvard.edu/abs/2018arXiv180706205P} {p. arXiv:1807.06205}

\bibitem[\protect\citeauthoryear{{Potter}, {Stadel}  \& {Teyssier}}{{Potter}
  et~al.}{2017}]{potter_etal2017}
{Potter} D.,  {Stadel} J.,   {Teyssier} R.,  2017, \mn@doi [Computational
  Astrophysics and Cosmology] {10.1186/s40668-017-0021-1}, \href
  {https://ui.adsabs.harvard.edu/abs/2017ComAC...4....2P} {4, 2}

\bibitem[\protect\citeauthoryear{{Prada}, {Klypin}, {Cuesta}, {Betancort-Rijo}
  \& {Primack}}{{Prada} et~al.}{2012}]{prada2012}
{Prada} F.,  {Klypin} A.~A.,  {Cuesta} A.~J.,  {Betancort-Rijo} J.~E.,
  {Primack} J.,  2012, \mn@doi [\mnras] {10.1111/j.1365-2966.2012.21007.x},
  \href {https://ui.adsabs.harvard.edu/abs/2012MNRAS.423.3018P} {423, 3018}

\bibitem[\protect\citeauthoryear{{Press} \& {Schechter}}{{Press} \&
  {Schechter}}{1974}]{pressschechter74}
{Press} W.~H.,  {Schechter} P.,  1974, \mn@doi [\apj] {10.1086/152650}, \href
  {https://ui.adsabs.harvard.edu/abs/1974ApJ...187..425P} {187, 425}

\bibitem[\protect\citeauthoryear{{Rey}, {Pontzen}  \& {Saintonge}}{{Rey}
  et~al.}{2019}]{Rey_etal2019}
{Rey} M.~P.,  {Pontzen} A.,   {Saintonge} A.,  2019, \mn@doi [\mnras]
  {10.1093/mnras/stz552}, \href
  {https://ui.adsabs.harvard.edu/abs/2019MNRAS.485.1906R} {485, 1906}

\bibitem[\protect\citeauthoryear{{Rodr{\'\i}guez-Puebla}, {Behroozi},
  {Primack}, {Klypin}, {Lee}  \& {Hellinger}}{{Rodr{\'\i}guez-Puebla}
  et~al.}{2016}]{bolplanck2016}
{Rodr{\'\i}guez-Puebla} A.,  {Behroozi} P.,  {Primack} J.,  {Klypin} A.,  {Lee}
  C.,   {Hellinger} D.,  2016, \mn@doi [\mnras] {10.1093/mnras/stw1705}, \href
  {https://ui.adsabs.harvard.edu/abs/2016MNRAS.462..893R} {462, 893}

\bibitem[\protect\citeauthoryear{{Roth}, {Pontzen}  \& {Peiris}}{{Roth}
  et~al.}{2016}]{Roth2016_GMhalo}
{Roth} N.,  {Pontzen} A.,   {Peiris} H.~V.,  2016, \mn@doi [\mnras]
  {10.1093/mnras/stv2375}, \href
  {https://ui.adsabs.harvard.edu/abs/2016MNRAS.455..974R} {455, 974}

\bibitem[\protect\citeauthoryear{{Salvador-Sol{\'e}}, {Solanes}  \&
  {Manrique}}{{Salvador-Sol{\'e}} et~al.}{1998}]{salvador-sole1998}
{Salvador-Sol{\'e}} E.,  {Solanes} J.~M.,   {Manrique} A.,  1998, \mn@doi
  [\apj] {10.1086/305661}, \href
  {https://ui.adsabs.harvard.edu/abs/1998ApJ...499..542S} {499, 542}

\bibitem[\protect\citeauthoryear{{Seljak}}{{Seljak}}{2000}]{seljak00}
{Seljak} U.,  2000, \mn@doi [\mnras] {10.1046/j.1365-8711.2000.03715.x}, \href
  {http://adsabs.harvard.edu/abs/2000MNRAS.318..203S} {318, 203}

\bibitem[\protect\citeauthoryear{{Skillman}, {Warren}, {Turk}, {Wechsler},
  {Holz}  \& {Sutter}}{{Skillman} et~al.}{2014}]{skillman2014}
{Skillman} S.~W.,  {Warren} M.~S.,  {Turk} M.~J.,  {Wechsler} R.~H.,  {Holz}
  D.~E.,   {Sutter} P.~M.,  2014, arXiv e-prints, \href
  {https://ui.adsabs.harvard.edu/abs/2014arXiv1407.2600S} {p. arXiv:1407.2600}

\bibitem[\protect\citeauthoryear{{Springel} et~al.,}{{Springel}
  et~al.}{2005}]{springel_etal05}
{Springel} V.,  et~al., 2005, \mn@doi [\nat] {10.1038/nature03597}, \href
  {http://adsabs.harvard.edu/abs/2005Natur.435..629S} {435, 629}

\bibitem[\protect\citeauthoryear{{Tasitsiomi}, {Kravtsov}, {Gottl{\"o}ber}  \&
  {Klypin}}{{Tasitsiomi} et~al.}{2004}]{tasitsiomi04}
{Tasitsiomi} A.,  {Kravtsov} A.~V.,  {Gottl{\"o}ber} S.,   {Klypin} A.~A.,
  2004, \mn@doi [\apj] {10.1086/383219}, \href
  {https://ui.adsabs.harvard.edu/abs/2004ApJ...607..125T} {607, 125}

\bibitem[\protect\citeauthoryear{{Wang} et~al.,}{{Wang}
  et~al.}{2011}]{wang2011}
{Wang} J.,  et~al., 2011, \mn@doi [\mnras] {10.1111/j.1365-2966.2011.18220.x},
  \href {https://ui.adsabs.harvard.edu/abs/2011MNRAS.413.1373W} {413, 1373}

\bibitem[\protect\citeauthoryear{{Warren}}{{Warren}}{2013}]{warren2013_2HOT}
{Warren} M.~S.,  2013, arXiv e-prints, \href
  {https://ui.adsabs.harvard.edu/abs/2013arXiv1310.4502W} {p. arXiv:1310.4502}

\bibitem[\protect\citeauthoryear{{Wechsler} \& {Tinker}}{{Wechsler} \&
  {Tinker}}{2018}]{wechsler_tinker18}
{Wechsler} R.~H.,  {Tinker} J.~L.,  2018, \mn@doi [\araa]
  {10.1146/annurev-astro-081817-051756}, \href
  {http://adsabs.harvard.edu/abs/2018ARA%26A..56..435W} {56, 435}

\bibitem[\protect\citeauthoryear{{Wechsler}, {Bullock}, {Primack}, {Kravtsov}
  \& {Dekel}}{{Wechsler} et~al.}{2002}]{wechsler02}
{Wechsler} R.~H.,  {Bullock} J.~S.,  {Primack} J.~R.,  {Kravtsov} A.~V.,
  {Dekel} A.,  2002, \mn@doi [\apj] {10.1086/338765}, \href
  {https://ui.adsabs.harvard.edu/abs/2002ApJ...568...52W} {568, 52}

\bibitem[\protect\citeauthoryear{{Wechsler}, {Zentner}, {Bullock}, {Kravtsov}
  \& {Allgood}}{{Wechsler} et~al.}{2006}]{wechsler06}
{Wechsler} R.~H.,  {Zentner} A.~R.,  {Bullock} J.~S.,  {Kravtsov} A.~V.,
  {Allgood} B.,  2006, \mn@doi [\apj] {10.1086/507120}, \href
  {https://ui.adsabs.harvard.edu/abs/2006ApJ...652...71W} {652, 71}

\bibitem[\protect\citeauthoryear{{White} \& {Rees}}{{White} \&
  {Rees}}{1978}]{whiterees78}
{White} S.~D.~M.,  {Rees} M.~J.,  1978, \mn@doi [\mnras]
  {10.1093/mnras/183.3.341}, \href
  {https://ui.adsabs.harvard.edu/abs/1978MNRAS.183..341W} {183, 341}

\bibitem[\protect\citeauthoryear{{Wu}, {Hahn}, {Wechsler}, {Mao}  \&
  {Behroozi}}{{Wu} et~al.}{2013}]{Wu2013}
{Wu} H.-Y.,  {Hahn} O.,  {Wechsler} R.~H.,  {Mao} Y.-Y.,   {Behroozi} P.~S.,
  2013, \mn@doi [\apj] {10.1088/0004-637X/763/2/70}, \href
  {https://ui.adsabs.harvard.edu/abs/2013ApJ...763...70W} {763, 70}

\bibitem[\protect\citeauthoryear{{Zhao}, {Mo}, {Jing}  \& {B{\"o}rner}}{{Zhao}
  et~al.}{2003}]{zhao03}
{Zhao} D.~H.,  {Mo} H.~J.,  {Jing} Y.~P.,   {B{\"o}rner} G.,  2003, \mn@doi
  [\mnras] {10.1046/j.1365-8711.2003.06135.x}, \href
  {https://ui.adsabs.harvard.edu/abs/2003MNRAS.339...12Z} {339, 12}

\bibitem[\protect\citeauthoryear{{Zhao}, {Jing}, {Mo}  \& {B{\"o}rner}}{{Zhao}
  et~al.}{2009}]{zhao_etal09}
{Zhao} D.~H.,  {Jing} Y.~P.,  {Mo} H.~J.,   {B{\"o}rner} G.,  2009, \mn@doi
  [\apj] {10.1088/0004-637X/707/1/354}, \href
  {https://ui.adsabs.harvard.edu/abs/2009ApJ...707..354Z} {707, 354}

\bibitem[\protect\citeauthoryear{{van den Bosch}}{{van den
  Bosch}}{2002}]{vdb02}
{van den Bosch} F.~C.,  2002, \mn@doi [\mnras]
  {10.1046/j.1365-8711.2002.05171.x}, \href
  {https://ui.adsabs.harvard.edu/abs/2002MNRAS.331...98V} {331, 98}

\bibitem[\protect\citeauthoryear{{van den Bosch}}{{van den
  Bosch}}{2017}]{vdBosch17}
{van den Bosch} F.~C.,  2017, \mn@doi [\mnras] {10.1093/mnras/stx520}, \href
  {https://ui.adsabs.harvard.edu/abs/2017MNRAS.468..885V} {468, 885}

\bibitem[\protect\citeauthoryear{{van den Bosch}, {Jiang}, {Hearin},
  {Campbell}, {Watson}  \& {Padmanabhan}}{{van den Bosch}
  et~al.}{2014}]{vdBosch14}
{van den Bosch} F.~C.,  {Jiang} F.,  {Hearin} A.,  {Campbell} D.,  {Watson} D.,
    {Padmanabhan} N.,  2014, \mn@doi [\mnras] {10.1093/mnras/stu1872}, \href
  {https://ui.adsabs.harvard.edu/abs/2014MNRAS.445.1713V} {445, 1713}

\bibitem[\protect\citeauthoryear{van~der Walt, Colbert  \& Varoquaux}{van~der
  Walt et~al.}{2011}]{numpy}
van~der Walt S.,  Colbert S.~C.,   Varoquaux G.,  2011, \mn@doi [Computing in
  Science Engineering] {10.1109/MCSE.2011.37}, 13, 22

\makeatother
\end{thebibliography}

\appendix

\section{Concentration--Mass Relation}
\label{sec:c-M_relation}

\begin{figure}
    \centering
    \includegraphics[scale=0.45]{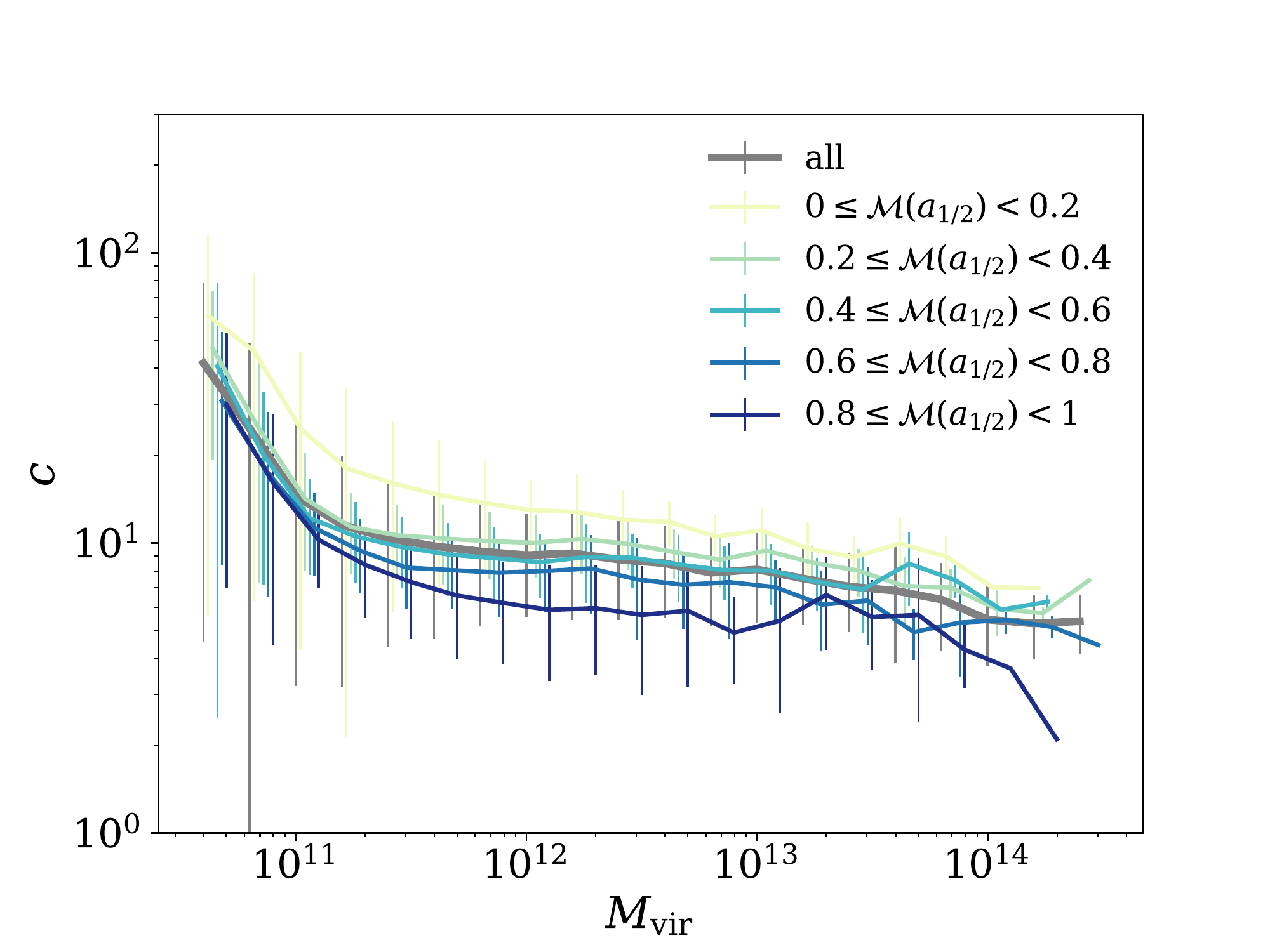}
    \caption{Concentration--mass relation for halo samples.
    The solid lines show the mean relations, while the error bars show the standard deviation of the relation within each sample.
    The thicker grey line shows the concentration--mass relation for all the haloes in the random sample described in \autoref{sec:amm_sample}, while the other lines show subsamples with different mark values of $a_{1/2}$, as are labelled in the legend.}
    \label{fig:c-M}
\end{figure}

In \autoref{fig:c-M}, we show the concentration--mass relation of the random sample described in \autoref{sec:amm_sample}, and the dependence of the relation on half-mass scales.
The random sample is divided into 5 quintiles based on the mass normalised marks of the half-mass scales.
In the figure, we show the mean concentration relation along with the standard deviation in the relation for each subsample.
The standard error of the mean is naturally much smaller.

From the figure we observe that, in general, the concentration--mass relation depends monotonically on the half-mass scale, again in consistence with previous findings that earlier forming haloes tend to be more concentrated.
The scatter in the relation is larger for later forming haloes except in a few cases, which can be explained by the fact that they are more likely to have had recent mergers.
This figure complements our findings in \autoref{fig:sigmalogc}.

\label{lastpage}
\end{document}